\documentclass[prc,showpacs,twocolumn,groupedaddress,superscriptaddress]{revtex4}
\usepackage{graphicx}
\usepackage{dcolumn}
\usepackage{bm}

\begin{document}

\title{Microscopic description of fission in  superheavy nuclei 
with the  parametrization D1M$^{*}$ of the Gogny energy density functional}

\author{R. Rodr\'{\i}guez-Guzm\'an}
\author{Y. M. Humadi}
\affiliation{ 
Physics Department, Kuwait University, Kuwait 13060, Kuwait.
}
\author{L. M. Robledo}
\affiliation{Center for Computational Simulation,
Universidad Polit\'ecnica de Madrid,
Campus de Montegancedo, Boadilla del Monte, 28660-Madrid, Spain
}
\affiliation{
Departamento  de F\'{\i}sica Te\'orica, 
Universidad Aut\'onoma de Madrid, 28049-Madrid, Spain}

\date{\today}

\begin{abstract}
The constrained Hartree-Fock-Bogoliubov 
approximation, based on the recent parametrization D1M$^{*}$ 
of the Gogny energy density functional, is used to describe fission in 435
superheavy nuclei. The Gogny-D1M$^{*}$ parametrization is benchmarked against available
experimental data on inner and  second barrier heights, excitation 
energies of the fission isomers and half-lives in a selected set of 
 Pu, Cm, Cf, Fm, No, Rf, Sg, Hs and Fl nuclei. Results are also compared 
with those obtained with the Gogny-D1M 
energy density functional. A detailed study of the minimal energy
fission paths is carried out for isotopic chains with atomic numbers
100 $\le$ Z $\le$ 126 including very neutron-rich sectors up to around 
4 MeV from the two-neutron driplines. Single-particle energies, ground state 
deformations, pairing correlations, two-nucleon separation energies 
and barrier heights 
are also discussed. In addition to fission paths, the 
constrained Hartree-Fock-Bogoliubov framework provides collective masses and zero-point  
quantum rotational and vibrational 
energies. Those quantities are  
building blocks within the Wentzel-Kramer-Brillouin formalism
employed to evaluate 
the systematic of the 
spontaneous fission half-lives t$_\mathrm{SF}$. The competition 
between spontaneous fission and $\alpha$-decay is studied, through the computation of
the $\alpha$-decay half-lives t$_\mathrm{\alpha}$ using a parametrization of the 
Viola-Seaborg formula. From the comparison with the available experimental data 
and the results obtained with other theoretical approaches, it is concluded 
that D1M$^{*}$ represents a reasonable starting point to describe fission in heavy 
and superheavy nuclei. 
\end{abstract}


\maketitle{}

%
%
%

\section{Introduction.}
The theoretical description of fission, one of the many possible decay 
modes of heavy atomic nuclei \cite{Specht,Bjor}, still remains one of 
the hottest topics at the frontier of contemporary nuclear physics.  
From a semi-classical perspective, fission can be viewed as the result 
of the competition between the Coulomb repulsion of the nuclear charge 
density and the nuclear surface energy 
\cite{MeitnerFrisch,Bohr-Wheeler-fission,proportional-1}. However, the 
evolution from the initial configuration (usually the ground state) to 
the scission point, where the nucleus starts to split into two or more 
fragments, strongly depends on the intermediate configurations whose 
properties are deeply influenced by subtle quantum mechanical effects 
associated with the evolution of shells with deformation, i.e.,  shell
effects. A detailed knowledge of those shell effects is 
required, for example, to better understand the very limits of the 
nuclear stability against fission. As one goes up in atomic number Z, 
Coulomb repulsion among protons overcomes the surface tension and  
quantum shell effects provide the only mechanism to increase the 
chances of survival  of a given element. Superheavy nuclei belong to 
the class of nuclear systems where there is no classical fission 
barrier and, therefore, only quantum shell effects are responsible for 
their stability. Therefore, they represent the perfect laboratory to 
understand the subtle quantum effects relevant for the fission dynamics 
\cite{Baran-NPA2015,Schunck2016} and to obtain key information on the existence of 
nuclei beyond Oganesson \cite{Naza-Nat-2018}.

%
%
\begin{figure*}
\includegraphics[width=1.0\textwidth]{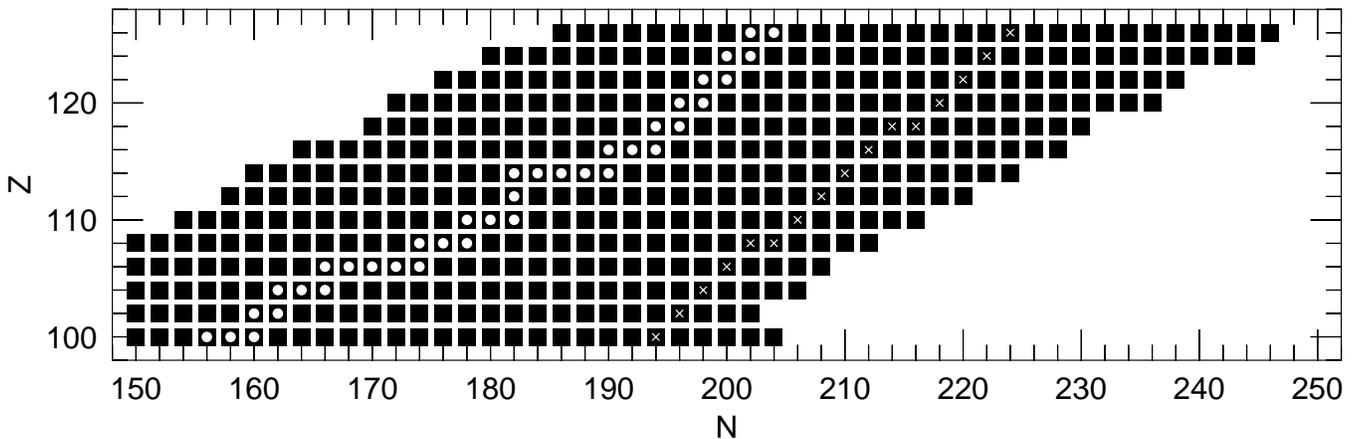}
\caption{The black squares represent the set of nuclei considered in this work. The white circles represent
the stability line obtained by looking at the minimum of the binding energy as a function of $N$ 
for constant values of $A$. The white crosses represent a typical r-process path that is drawn here
with the sole purpose of giving an idea of the relevance of the present study for r-process
calculations. On the neutron deficient side we reach the proton drip line (Z $>$ 106) and in the
neutron rich side the frontier corresponds to $S_{2n} \sim 4 \textrm{MeV}$.
}
\label{studied_nuclei} 
\end{figure*}

One of the key applications of nuclear fission concerns the r-process 
nucleosynthesis of superheavy elements. In scenarios like the dynamical 
ejecta of neutron start mergers, the competition between rapid neutron 
captures and beta decays of seed nuclei leads to the synthesis of 
superheavy elements. The r-process path proceeds to regions of unstable 
nuclei that undergo fission, recycling the material to lighter  
products \cite{Goriely-app1,Cowan-app1,MP-app1}. Therefore, fission 
plays a key role to modify the final shape of the r-process abundances 
\cite{Abunda-1,Abunda-2} as well as to achieve a robust r-process 
\cite{Robust-Mendoza}. Systematic studies of the fission paths and 
related properties are also very useful to deepen our knowledge of  the 
competition between different decay channels (fission, $\alpha$-decay, 
$\dots$) in heavy and superheavy nuclei 
\cite{Viola-Seaborg,TDong2005,Rayner-fission-1,Rayner-fission-2,Rayner-fission-3,Rayner-fission-4,Rayner-fission-5,Robledo-Giulliani,Min_Action_Gogny,Min_Action_GognyRayner2018}. 
In addition, nuclear fission remains a topic of high interest for 
reactor physics, the degradation of radioactive waste, prompt neutron 
capture data from weapon tests as well as  in the context of the 
synthesis of superheavy elements  in laboratories all over around the 
world 
\cite{Specht,Krappe,Wagemans,Sierk-PRC2015,JULIN-SHE,Oganessian-3,Haba-SHE}.

From a theoretical point of view, the macroscopic-microscopic (Mac-Mic)
model 
\cite{Strutisnky-ShellCorr-Method-1,Strutisnky-ShellCorr-Method-2,proportional-1,Myers-Swiatecki,Moller-1,Moller-2,PatykN152N162} has already been successfully applied in systematic studies 
of superheavy nuclei. Such an approach, provides a very accurate description of the fission
landscape in terms of up to five collective degrees of freedom  \cite{Sierk-PRC2015,Jachi-SHE-1,Jachi-SHE-2}. In recent years 
the constrained mean-field approximation \cite{rs} has 
emerged as a useful tool 
for microscopic fission studies \cite{Schunck2016,Baran-NPA2015}. Calculations are typically 
carried out with  non-relativistic  Gogny 
\cite{gogny-d1s,Berger.01,Delaroche-2006,Robledo-Martin,Dubray,PEREZ-ROBLEDO,Younes2009,WERP02,Egido-other1, 
Warda-Egido-2012}, Skyrme 
\cite{UNEDF1,Mcdonell-2,Erler2012,Baran-SF-2012,Baran-1981}, and 
Barcelona-Catania-Paris-Madrid (BCPM) 
\cite{Robledo-Giulliani,Giuliani-Pinedo-SHE-rp} as well as  with relativistic 
\cite{Bender-1998,Abusara-2010,Abu-2012-bheights,RMF-LU-2012,Kara-RMF,Agbemava-Hyperheavy-2019,vanishing-Agbe-2017-RMF} 
energy density functionals (EDFs). The (constrained) mean-field approximation 
provides selfconsistent potential energy surfaces as functions of the 
quadrupole, octupole, $\dots$ deformations. Quantum shell effects are  included from the beginning in this
unique framework.
Moreover, the mean-field approximation also provides in a consistent framework the 
associated collective inertias as well as the quantum zero-point  
rotational and vibrational energy corrections. All those are basic ingredients 
required to compute fission observables like, for example, the 
spontaneous fission  half-lives t$_\mathrm{SF}$ 
\cite{Schunck2016,Baran-NPA2015}.

\begin{table*}
\label{Table1}
\caption{The heights of the inner  $B_{I}^{th}$ and second $B_{II}^{th}$
 barriers as well as the excitation energies $E_{II}^{th}$ of the fission 
isomers, predicted with the Gogny-D1M$^{*}$ EDF, are compared with the available 
experimental values  $B_{I}^{exp}$, $B_{II}^{exp}$ and $E_{II}^{exp}$ 
\cite{Refs-barriers-other-nuclei-1,Refs-barriers-other-nuclei-2}. 
The $B_{I}^{th}$ values obtained considering triaxial shapes
are given in parenthesis.
Results obtained with the Gogny-D1M EDF \cite{Rayner-fission-1}
are  included in the table. For each nucleus, the value of $Z^{2}$/A
is also included in the table to facilitate the comparison of our results with other models/calculations.
}
\begin{tabular}{ccccccccccc}
\hline
\hline
Nucleus  & $Z^{2}/A$~  &  $B_{I}^{th}(D1M)$ & $B_{I}^{th}$(D1M$^{*}$)   &  $B_{I}^{exp}$     & $E_{II}^{th}(D1M)$ & $E_{II}^{th}$(D1M$^{*}$) &  $E_{II}^{exp}$ & $B_{II}^{th}(D1M)$  & $B_{II}^{th}$(D1M$^{*}$) &  $B_{II}^{exp}$  \\
\hline
\hline
$^{234}$U& 36.17 &   7.60		&     7.47			 &  4.80	&  3.32 	     &    2.80  		&      -	   &  
8.09	    &   7.72             &      5.50  	  \\ 
         &  &  (7.01)		&     (6.79)			 &		&		     &  			&		   &  
	   &			&	\\ 	  	   
\hline
$^{236}$U& 35.86 &  8.33		&      8.34			  &  5.00	 &  3.17	      &    2.67 		 &     2.75	     &  
8.69	    &  8.07  	         &      5.67               \\
         &  &  (7.00)		&      (7.15)			 &		 &		      & 			 &		     &  
 	   &			&			   \\
\hline
$^{238}$U& 35.56  &  9.06		 &	9.14			   &  6.30	  &  3.37	       &    2.82		  &	2.55	      & 
9.54	    &  8.97              &      5.50  	       \\
         &   &  (7.46)		 &	(7.79)  		   &		  &		       &			  &		      & 
 	    &			 &	\\	       	 
\hline
$^{238}$Pu&37.13  &  8.77		   &	  8.83  		     &  5.60	    &  3.20		 &    2.72		    &	  2.40  	& 
7.75	    &  7.20              &      5.10  	       \\
          &  &  (7.66)		 &	(7.80)  		   &		  &		       &			  &		      & 
	    &			 &		       \\
\hline 
$^{240}$Pu& 36.82 &  9.45		 &	9.56			   &  6.05	  &  3.36	       &    2.78		  &	2.80	      &  
8.57	    &  8.00              &      5.15  	       \\
          &  &  (7.70)		 &	(7.81)  		   &		  &		       &			  &		      &  
	    &			 &		       \\
\hline 
$^{242}$Pu& 36.51 &  9.90		 &	10.08			   &  5.85	  & 3.57	       &    2.90		  &	2.20	      & 
9.18	    &  8.66              &      5.05  	       \\
          &  &  (7.67)		 &	(8.12)  			 &	  &		       &			  &		      & 
 	    &			 &		       \\
\hline 
$^{244}$Pu& 36.21 &  10.16		 &	10.32			   &  5.70	  & 3.83	       &    3.15		  &	-	      &  
9.60	    &  9.03              &      4.85  	       \\
          &  &  (7.42)		 &	(7.55)  			 &	  &		       &    3.15		  &		      &  
 	    &			 &		       \\
\hline
$^{240}$Cm& 38.40 &  8.98		 &	8.91			   &  - 	  & 2.55	       &    2.19		  &	2.00	      & 
6.13	    &  5.79                  &  -		    \\
          &  &  (7.87)		 &	(7.83)  		   &		  &		       &			  &		      & 
 	    &			 &		    \\
\hline 
$^{242}$Cm& 38.08 &  9.78		 &	9.69			   &  6.65	  & 2.77	       &   2.31 		  &	1.90	      & 
6.99	    &  6.53              &       5.00  	       \\
          &  &  (8.31)		 &	(8.13)  		   &		  &		       &			  &		      & 
	    &			 &		       \\
\hline 
$^{244}$Cm& 37.77 & 10.38		 &	11.02			   &  6.18	  & 3.02	       &   2.47 		  &	2.20	      & 
7.70	    &  7.15              &       5.10  	        \\
          &  &  (8.27)		 &	(9.65)  		   &		  &		       &			  &		      & 
 	    &			 &			\\
\hline
$^{246}$Cm& 37.46 &  10.75		 &	10.87			   &  6.00	  & 3.29	       &   2.85 		  &	-	      & 
8.13	    & 7.89               &      4.80  	       \\
          &  &  (8.03)		 &	(8.26)  		   &		  &		       &			  &		      & 
 	    &			 &		       \\
\hline 
$^{248}$Cm& 37.16 & 10.68		 &	9.82			   &  5.80	  & 3.32	       &    2.93		  &	-	      & 
8.28	    & 8.00               &      4.80  	       \\
          &  & (7.50)		 &	(7.53)  		   &		  &		       &			  &		      & 
 	    &			 &		       \\
\hline 
$^{250}$Cf& 38.42 &  11.38		 &	10.65			   &  - 	  & 2.81	       &    2.58		  &	-	      & 
7.09	    &  6.99              &      3.80  	       \\
          &  &  (8.25)		 &	(7.74)  		   &		  &		       &			  &		      & 
 	    &			 &		       \\
\hline 
$^{252}$Cf& 38.11 &  10.96		 &	10.39			   &  - 	  & 1.37	       &    2.18		  &	-	      & 
6.79	    &  6.63  	         &       3.50           \\
          &  &  (8.07)		 &	(7.27)  		   &		  &		       &			  &		      & 
 	    &			 &		     \\
\hline  
\hline 
\end{tabular}
\end{table*}

The mean-field framework  implicitly assumes, that 
fission properties are determined by general features of the employed
EDF. Among the members of the D1 family of  
parametrizations of the Gogny-EDF, D1S
\cite{gogny-d1s} 
has already built a strong reputation given its ability to reproduce 
a wealth of low-energy nuclear data all over the nuclear chart  both at the mean-field level and beyond
\cite{Review}.
In particular, the Gogny-D1S EDF has already been  applied to  study 
heavy and superheavy nuclei \cite{Berger.01,Delaroche-2006,WERP02,Warda-Egido-2012}.
However, one of the deficiencies of the parametrization D1S is the drifting in binding 
energies along isotopic chains associated to the not so satisfactory neutron matter
equation of state (EoS), as compared to more realistic calculations \cite{EoS-VP}.
To cure this deficiency the 
D1N parametrization \cite{gogny-d1n} was introduced. This parametrization, however, has 
scarcely been  used in the literature 
\cite{Rayner-fission-1,PTpaper-Rayner,Robledo-Bertsch-systQ30,RobledoJPG-Q302015,Robledo-Rayner-JPG-2012}. 
On the other hand, the parametrization D1M \cite{gogny-d1m} included in its fitting protocol not only realistic neutron matter
EoS information but also the binding energy of all known nuclei. With an impressive rms for binding 
energies of 0.798 MeV, it represents an excellent and competitive choice to deal with 
nuclear masses. The suitability of the Gogny-D1M EDF to describe fission in heavy 
and superheavy nuclei has been demonstrated in previous studies 
\cite{Rayner-fission-1,Rayner-fission-2,Rayner-fission-3,Rayner-fission-4,Rayner-fission-5}
where the results for barrier heights, excitation energies of fission isomers and half-lives 
have shown a reasonable  agreement with the available experimental data
as well as with other theoretical studies using different interactions
\cite{Delaroche-2006,Robledo-Giulliani,WERP02,Warda-Egido-2012}. In addition, D1M 
essentially retains the same predictive power as D1S 
in the description of 
nuclear structure phenomena 
\cite{PRCQ2Q3-2012,PTpaper-Rayner,Rayner-Sara,Rayner-Robledo-JPG-2009,Robledo-Rayner-JPG-2012,Rayner-PRC-2010,Rayner-PRC-2011}.

Unfortunately, the physics of neutron stars (NSs) is not well reproduced
by any of the Gogny force parametrizations introduced so far. All the
parametrizations turn out to be unable 
\cite{unable-1,unable-2,unable-3} to provide NS masses of about two 
solar masses 2\(M_\odot\), as required by recent astrophysical 
observations \cite{astrophy-obs-1,astrophy-obs-2}. Moreover, only the 
Gogny-D1M EDF achieves a NS mass above the canonical value 
1.4\(M_\odot\). In order to deal with this problem, a new 
reparametrization of the Gogny-D1M EDF has been proposed recently. In 
the new parametrization, denoted D1M$^{*}$ \cite{gogny-D1MSTAR}, the 
slope $L$ of the symmetry energy coefficient in nuclear matter is 
fitted to a value (43.18 MeV) larger than the one in D1M (24.83 MeV) 
and more according to the expected value for this coefficient. This 
modification of the parameter $L$ leads to a stiffer EOS for $\beta$-stable 
matter and to NS masses of 2\(M_\odot\). The reparametrization is
carried out in a way that all the other relevant combinations of 
parameters keep their values as to preserve most of the properties of the 
Gogny-D1M EDF \cite{gogny-d1m,gogny-D1MSTAR}. It has been checked, by 
means of selected calculations, that D1M$^{*}$ is as good as D1M in 
describing properties of finite nuclei \cite{D1MSTARFN}. However, much more work is 
still required to assess the performance of D1M$^{*}$, especially in 
the case of fission.

\begin{figure*}
\includegraphics[width=1.0\textwidth]{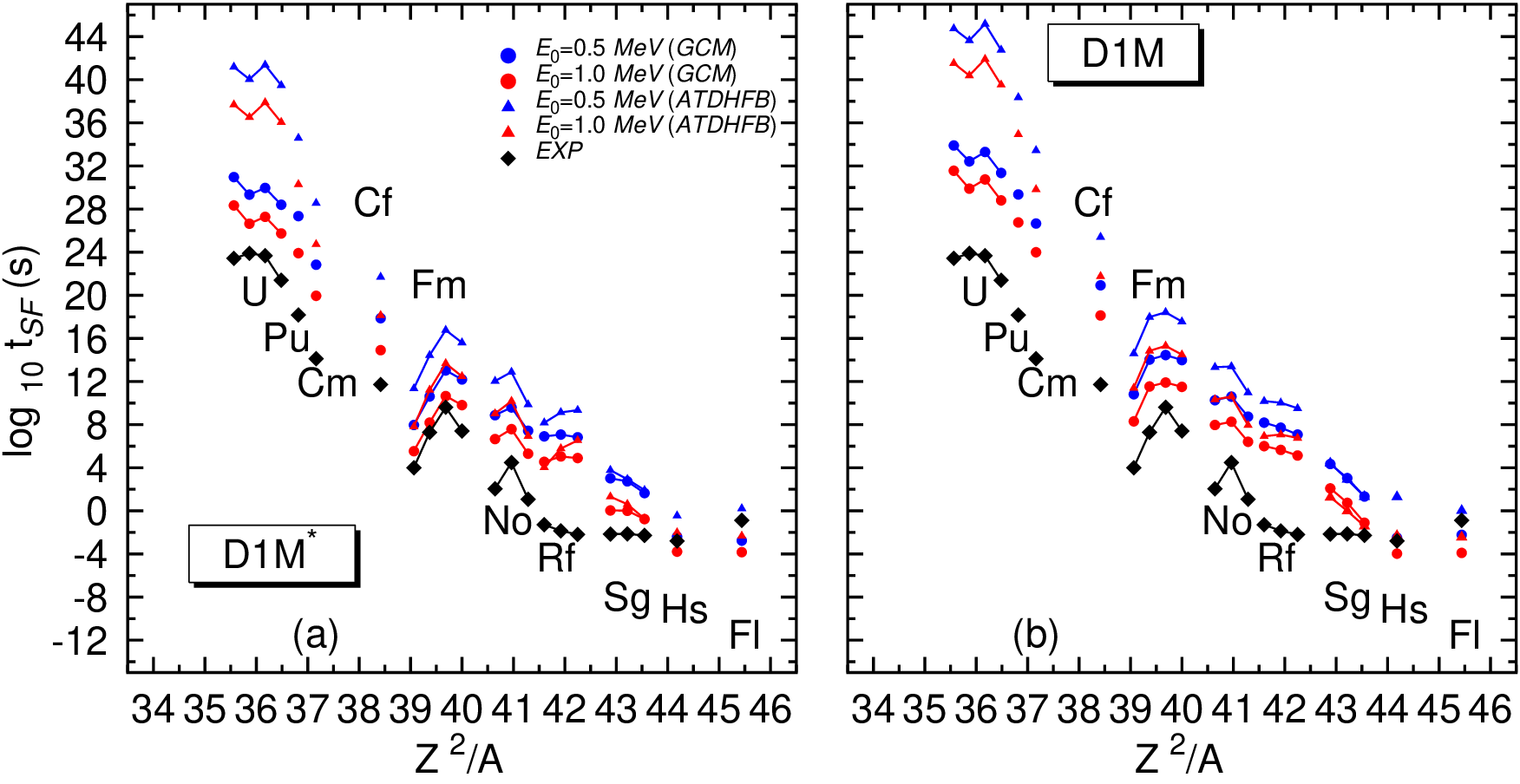}
\caption{(Color online) The spontaneous fission half-lives t$_\mathrm{SF}$ obtained for the nuclei
$^{232-238}$U, $^{240}$Pu, $^{248}$Cm, $^{250}$Cf, $^{250-256}$Fm, $^{252-256}$No,
$^{256-260}$Rf, $^{258-262}$Sg, $^{264}$Hs and $^{286}$Fl  within the GCM and ATDHFB schemes 
are depicted, as functions of the fissility-related parameter $Z^{2}$/A, in panel (a) for the 
Gogny-D1M$^{*}$ and in panel (b) for the Gogny-D1M EDFs. They are compared with the 
corresponding 
experimental data \cite{Refs-barriers-other-nuclei-3-tsf,Bertsch15}. Theoretical results 
are shown for E$_{0}$ = 0.5 MeV and E$_{0}$ =1.0 MeV. For more details, see the main text.
}
\label{tsf-nuclei-with-data} 
\end{figure*}

The improved value of $L$ leads to the expectation that the nuclear 
properties of very neutron-rich nuclei obtained with D1M$^{*}$ are 
going to be more realistic than the ones obtained with D1M. As a 
consequence, D1M$^{*}$ seems to be the appropriate choice to obtain 
realistic fission properties of very neutron-rich superheavy nuclei as 
required by astrophysical simulations of nucleosynthesis in the 
two-neutron start merger scenario. Very neutron-rich superheavy elements live in an 
unexplored territory where there is an enormous deficit of experimental 
information and therefore microscopic nuclear structure input is still 
much required \cite{Giuliani-Pinedo-SHE-rp,required}. In this study we 
use the Hartree-Fock-Bogoliubov (HFB) mean field scheme \cite{rs} along 
with the D1M$^{*}$ parametrization to describe the fission properties 
of the 435 even-even superheavy systems shown in 
Fig.~\ref{studied_nuclei}. The set of nuclei correspond, for each $Z$ value, to roughly 10 nuclei to the left
(with respect to the N axis) and 20 to the right with respect to the stability line. This corresponds, in
the neutron deficient side to almost the proton drip line. In the neutron rich side we go up to nuclei with $S_{2n} \sim 4 \textrm{MeV}$ covering a 
wide region of nuclei where the r-process takes place. The suitability of the 
Gogny-D1M$^{*}$ HFB framework to capture basic fission properties is confirmed by 
comparing our results for another 22 additional nuclei (see Table \ref{Table1} and Fig \ref{tsf-nuclei-with-data}) with available experimental data
\cite{Refs-barriers-other-nuclei-1,Refs-barriers-other-nuclei-2,Refs-barriers-other-nuclei-3-tsf,Bertsch15} .

The paper is organized as follows. In Sec. ~\ref{Theory-used}, we 
briefly outline the theoretical framework used in this study. The 
results of our calculations are discussed in Sec.~\ref{RESULTS}. First, 
in Sec.~\ref{validate}, we discuss the D1M$^{*}$ results for the nuclei 
$^{232-238}$U, $^{238-244}$Pu, $^{240-248}$Cm, $^{250,252}$Cf, 
$^{250-256}$Fm, $^{252-256}$No, $^{256-260}$Rf, $^{258-262}$Sg, 
$^{264}$Hs and $^{286}$Fl and compare them with available experimental 
data 
\cite{Refs-barriers-other-nuclei-1,Refs-barriers-other-nuclei-2,Refs-barriers-other-nuclei-3-tsf,Bertsch15} 
as well as with results obtained with the Gogny-D1M EDF. In 
Sec.~\ref{methodology}, we illustrate the methodology employed to 
compute the fission paths and other fission-related quantities using 
the nuclei $^{256}$No and $^{298}$No as examples. Both sections, are 
mainly intended to validate the Gogny-D1M$^{*}$ EDF for fission 
studies. We end Sec.~\ref{methodology}, with a brief discussion of the 
neutron and proton single-particle energies, as functions of the 
quadrupole moment Q$_{20}$,  in the case of $^{266}$Rf which is taken 
as an illustrative example. The systematic of the fission paths and 
the  fission half-lives obtained for the even-even nuclei shown in 
Fig.~\ref{studied_nuclei} is presented in Sec.~\ref{FB-systematcis}. In 
the same section, we also pay attention to the competition between 
spontaneous fission and $\alpha$-decay. In addition, ground state 
deformations, pairing correlations, two-nucleon separation energies and 
barrier heights are discussed. Finally,  Sec.~\ref{conclusions} is 
devoted to the concluding remarks and work perspectives.

%
%

%
%
%

\section{Theoretical framework}
\label{Theory-used}
In this section, we briefly outline the theoretical framework used in 
the present  study. A detailed account of our methodology can be found 
in Refs.~\cite{Rayner-fission-1,Rayner-fission-2,Rayner-fission-3}. We 
have resorted to the HFB approximation with constrains on the axially 
symmetric quadrupole  
$\hat{Q}_{20}= z^{2} -\frac{1}{2} \Big(x^{2} + y^{2} \Big)$
and octupole 
$\hat{Q}_{30} = z^{3} -\frac{3}{2} z \Big(x^{2} + y^{2} \Big)$ 
operators \cite{PRCQ2Q3-2012,Robledo-Rayner-JPG-2012}. The quadrupole  
and octupole moments are then computed as   $Q_{20} = \langle 
\hat{Q}_{20} \rangle$ and  $Q_{30} = \langle \hat{Q}_{30} \rangle$. The 
corresponding 
deformation parameters $\beta_{2}$ and $\beta_{3}$  read
\begin{equation} \label{beta2def}
\beta_{2} = \sqrt{\frac{4 \pi}{5}} \frac{ Q_{20} }{\langle r^{2} \rangle}
\end{equation}
and 
\begin{equation} \label{beta3def}
\beta_{3} =  \frac{\sqrt{7 \pi} Q_{30}}{3 \langle r^{3} \rangle}
\end{equation}
where, $\langle \dots \rangle$ represents the average taken with the 
corresponding HFB wave function. Aside from these constrains, we 
include the constrains on both the proton and neutron numbers 
\cite{rs}, characteristic of the HFB method. Finally, a constrain on 
the operator $\hat{Q}_{10}$ is used 
\cite{PRCQ2Q3-2012,Robledo-Rayner-JPG-2012} to prevent spurious effects 
associated to the center of mass motion that might appear when 
reflection symmetry (parity) is broken.

The HFB quasiparticle creation and annihilation operators 
$\beta^{\dagger}_{\mu}$ and $\beta_{\mu}$ have been expanded in an 
axially symmetric (deformed) harmonic oscillator  (HO) basis containing  
states with $J_{z}$ quantum numbers up to 31/2 and up to 26 quanta in 
the z direction. The basis quantum numbers are restricted by the 
condition $2n_{\perp} + |m| + \frac{1}{q} n_{z}  \le  M_{z,MAX}$ with 
$M_{z,MAX}$ = 17 and $q$ = 1.5. This choice of the parameter $q$ is 
dictated by the elongated prolate shapes typical of the fission process 
and is the same used in other similar calculations 
\cite{Robledo-Giulliani,WERP02,Rayner-fission-1,Rayner-fission-2,Rayner-fission-3,Rayner-fission-4,Rayner-fission-5}. 
For each of the studied nuclei and each of the constrained 
configurations along the corresponding fission paths, the two lengths  
$b_{z}$ and $b_{\perp}$ characterizing the HO basis have been optimized 
so as to minimize the total HFB energy. The computationally expensive 
oscillator length optimization along with the large HO basis used 
guarantees good convergence in the relevant range of shape deformations 
considered in this work \cite{Rayner-fission-2,WERP02}.

For the solution of the HFB equations, an approximate second order 
gradient method  \cite{Gradient-2order} has been used. This method
guarantees a rapid convergence of the self-consistent HFB procedure as
well as an efficient handling of the many constrains imposed. 
As it is customary in all the parametrizations of the Gogny-EDF, the 
two-body kinetic energy correction, including the exchange and pairing 
channels, has been taken into account in the Ritz-variational 
procedure. On the other hand, the Coulomb exchange term is considered 
in the Slater approximation \cite{CoulombSlater} while the Coulomb and 
spin-orbit contributions to the pairing field have been neglected.

%
%
\begin{figure*}
\includegraphics[width=0.45\textwidth]{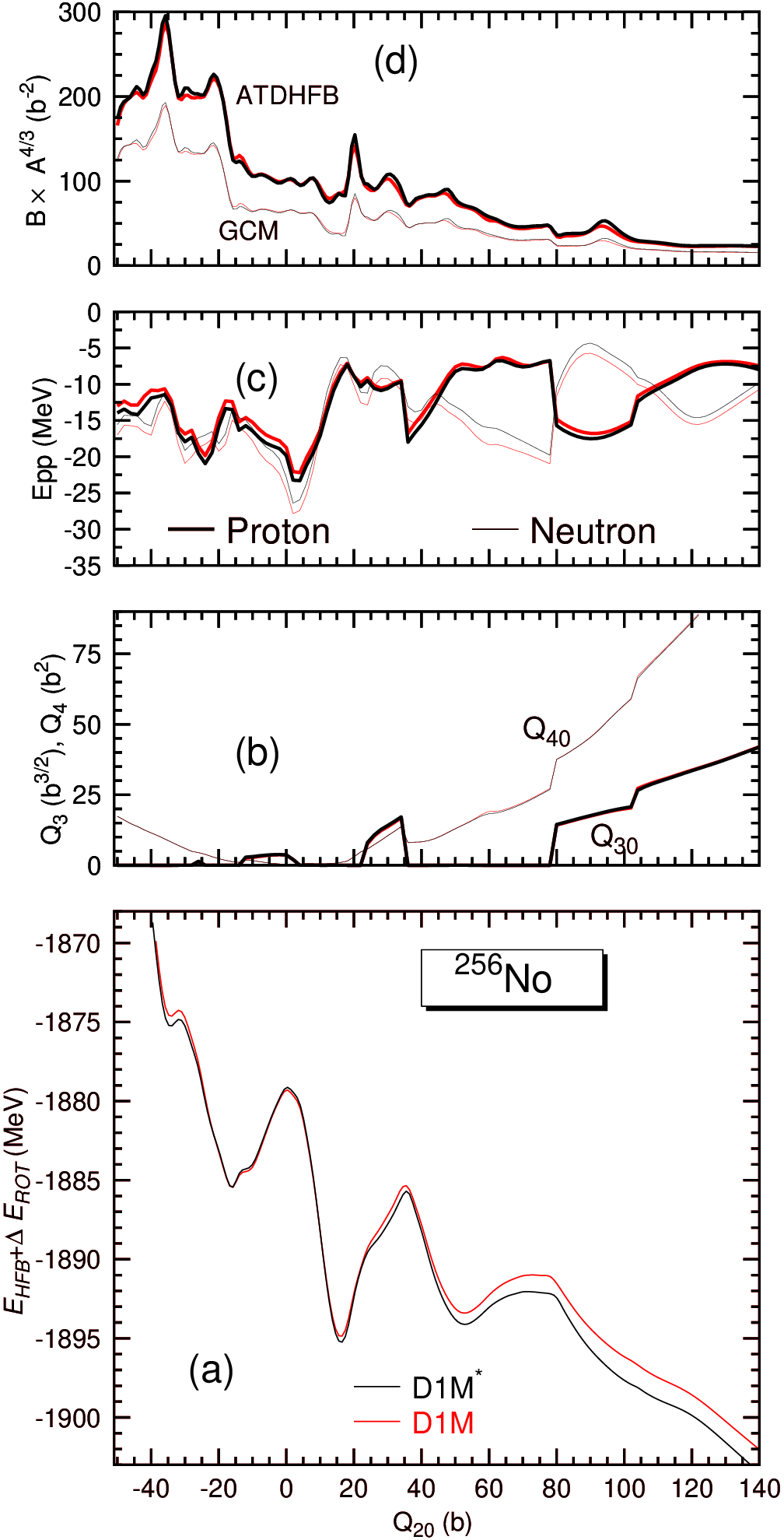}
\includegraphics[width=0.45\textwidth]{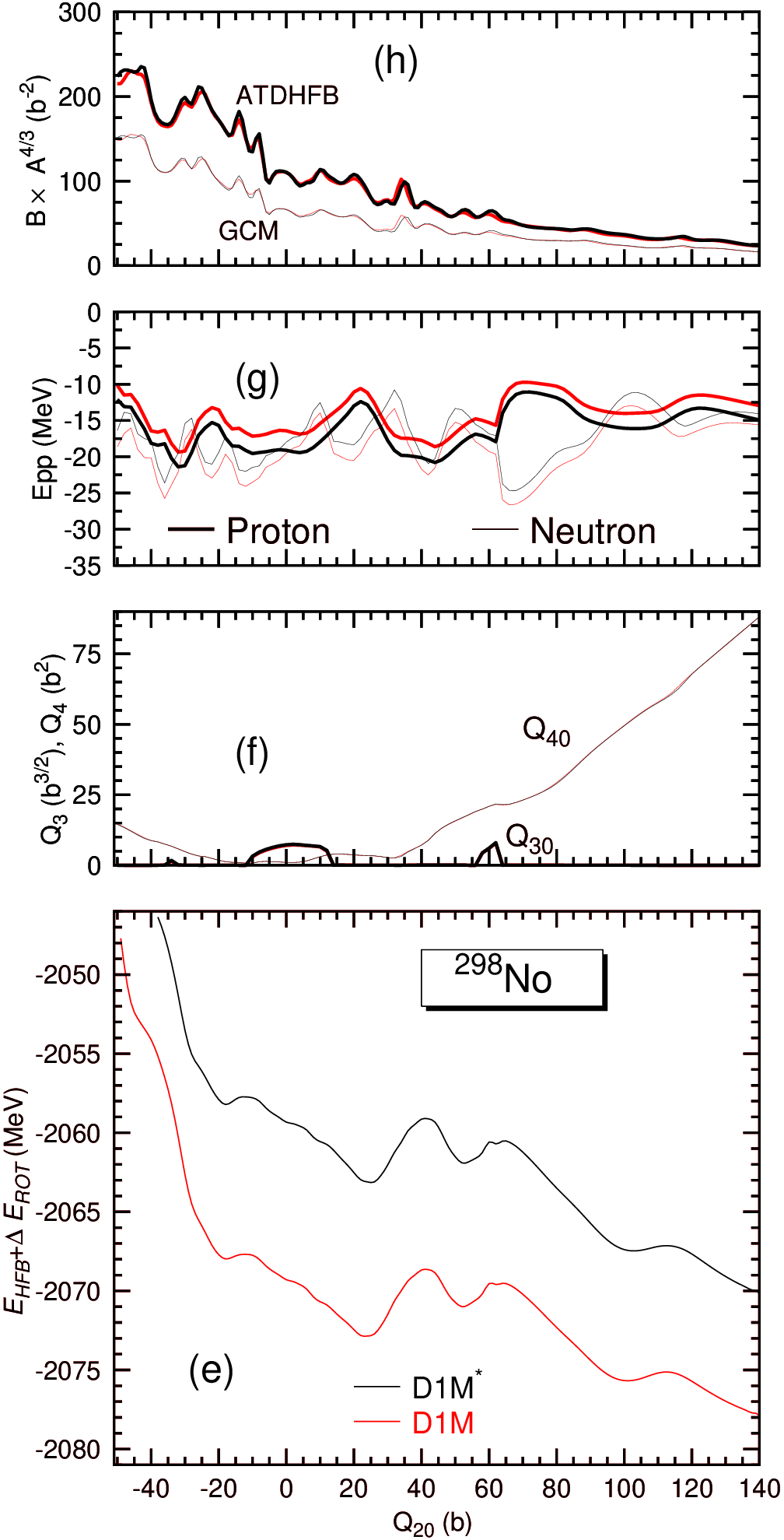}
\caption{(Color online) The HFB plus the zero point rotational energies 
obtained for the nucleus $^{256}$No ($^{298}$No) are plotted in panel 
(a) [panel (e)] as functions of the quadrupole moment Q$_{20}$. The 
octupole (thick lines) and hexadecupole (thin lines) moments are 
plotted in panel (b) [panel (f)]. The pairing interaction energies are 
depicted in panel (c) [panel (g)] for protons (thick lines) and 
neutrons (thin lines). The collective masses obtained within the ATDHFB 
(thick lines) and GCM (thin lines) schemes are plotted in panel (d) 
[panel (h)]. Results have been obtained with the parametrizations D1M 
(red curves) and D1M$^{*}$ (black curves) of the Gogny-EDF. For more 
details, see the main text. }
\label{FissionBarriers_pedagogical} 
\end{figure*}

We have computed the spontaneous fission half-life (in seconds) within 
the  Wentzel-Kramer-Brillouin (WKB) formalism 
\cite{proportional-1,Schunck2016} as 
\begin{equation} \label{TSF}
t_\mathrm{SF}= 2.86 \times 10^{-21} \times \left(1+ e^{2S} \right)
\end{equation}
where the action $S$ along the (minimal energy one-dimensional projected)
fission path reads
\begin{equation} \label{Action}
S= \int_{a}^{b} dQ_{20} \sqrt{2B(Q_{20})\left(V(Q_{20})-\left(E_\mathrm{min}+E_{0} \right)  \right)}
\end{equation}  
The integration limits $a$ and $b$ correspond to the classical turning 
points \cite{proportional-1} for the energy $E_\mathrm{min}+\mathrm{E}_{0}$. The energy 
$E_\mathrm{min}$ corresponds to the ground state minimum for the considered 
path while E$_{0}$ accounts for the ground state's quantal zero point energy associated to 
the quadrupole collective degree of freedom. In this work, 
we have performed calculations with two typical values E$_{0}$ = 0.5 
MeV and E$_{0}$ =1.0 MeV as it is customary in fission studies 
\cite{WERP02,Warda-Egido-2012,Rayner-fission-1}. Such an 
approximation ignores the dependence of E$_{0}$ on the 
softness of the collective potential as well as its dependence 
on particle number. 
However, in future more microscopic studies of the E$_0$ values are needed in
order to verify whether this conventional range of 0.5-1.0 MeV is accurate
enough.

In Eq.~(\ref{Action}), $B(Q_{20})$ represents the collective mass for the quadrupole
degree of freedom. The 
collective potential  $V(Q_{20})$ is given by the HFB energy minus the
zero-point rotational energy correction $\Delta E_{ROT}(Q_{20}) = \langle \vec{J}^{2}\rangle / (2 \mathcal{J}_{Y})$ and minus
the zero point energy associated to the axial quadrupole vibration $\Delta E_{VIB}(Q_{20})$. For the evaluation of 
the collective mass  and the zero-point vibrational energy correction  
two methods have been used. One is the ``perturbative" approximation 
\cite{crankingAPPROX,Giannoni,Libert-1999} to the Adiabatic Time 
Dependent HFB (ATDHFB) scheme. The second method is based on the 
Gaussian Overlap Approximation (GOA)  to the GCM and also using the
``perturbative" approximation. Details on how to 
compute the required  quantities can be found, for instance, in 
Refs.~\cite{Krappe,Schunck2016,Rayner-fission-1}. The  rotational 
correction $\Delta E_{ROT}(Q_{20})$ has been computed, in terms of the 
Yoccoz moment of inertia $\mathcal{J}_{Y}$, according to the phenomenological 
prescription discussed in Refs.~\cite{RRG23S,ER-Lectures}. In principle, 
the numerical pre-factor in front of the 
exponential of the action in  Eq. (\ref{TSF}) and related to the assault frequency, should be
computed taking into account the properties of the ground state minimum. However, this effect amounts at most to
a factor of 2 in t$_\mathrm{SF}$ which is much smaller than the 
uncertainties in the estimation of the half-life arising from 
the theoretical uncertainties of the ingredients of the action \cite{Rayner-fission-1}. 
Let us also mention that, in the 
computation of the spontaneous fission half-lives, the wiggles in the 
collective masses have been softened using a three point filter to 
facilitate the numerical evaluation of the required integrals 
\cite{Rayner-fission-1,Rayner-fission-2,Rayner-fission-3}. Those wiggles
are due to level crossings in the single particle spectrum. As the inertia
is computed using the ``perturbative" approximation the effect of level 
crossings is more intense than in the case of an exact evaluation of the 
inertia.

%
\begin{figure*}
\includegraphics[width=0.45\textwidth]{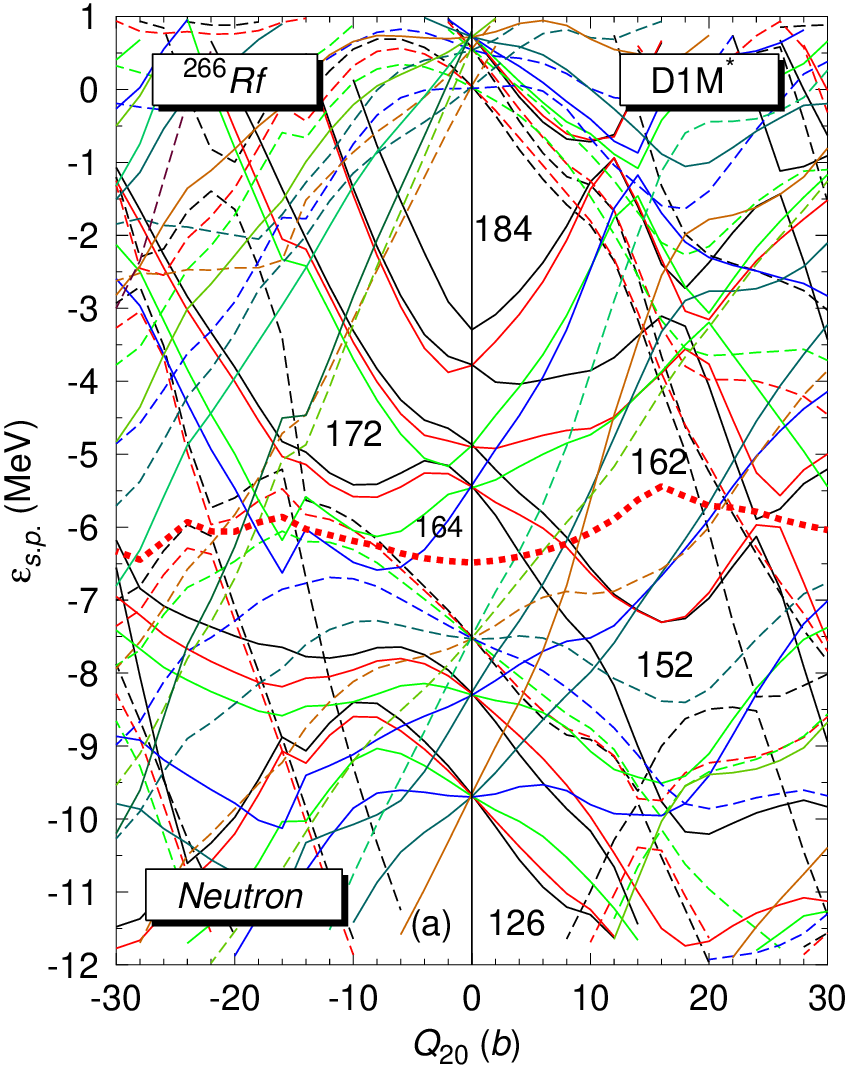}
\includegraphics[width=0.45\textwidth]{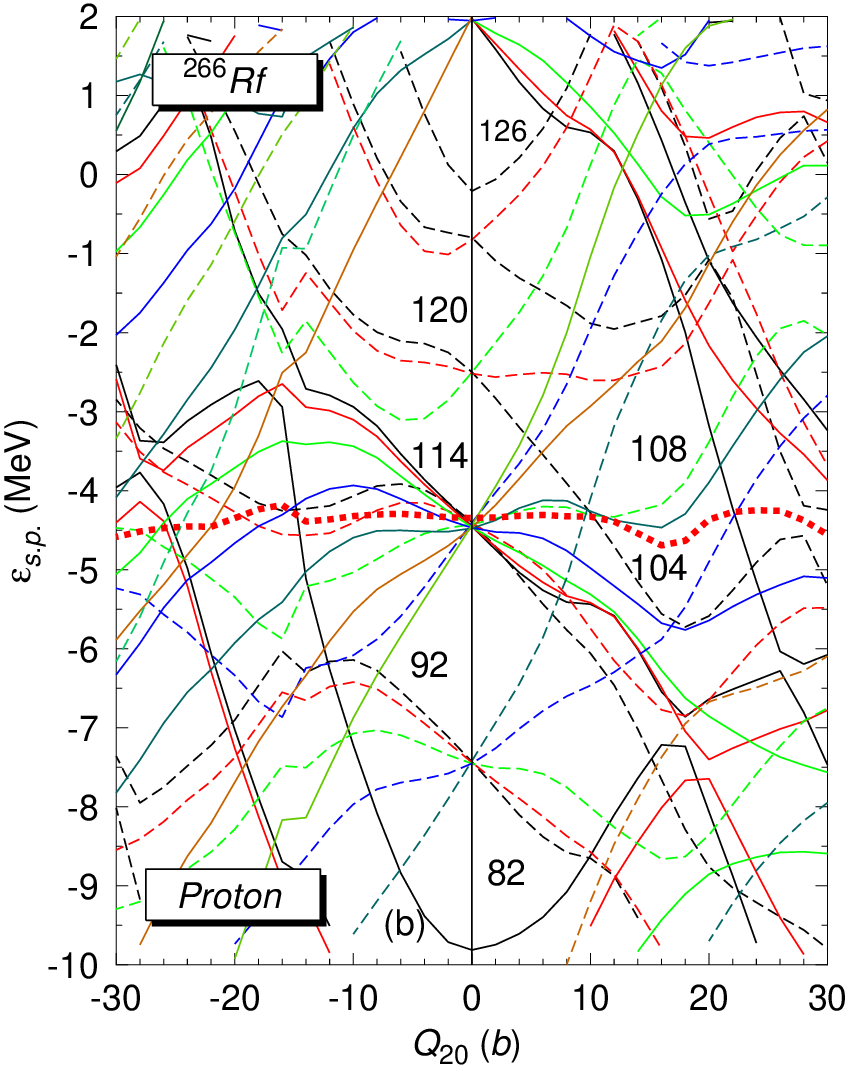}
\caption{(Color online) Neutron [panel (a)] and proton [panel (b)] 
single-particle energies, as functions of the quadrupole moment 
Q$_{20}$, for the nucleus $^{266}$Rf. The Fermi levels are also plotted 
with a thick (red) dotted line. Results correspond to the 
Gogny-D1M$^{*}$ EDF. Solid (dashed) lines are used for positive 
(negative) parity states. With increasing K =1/2, 3/2, 5/2, $\dots$ 
values color labels are black, red, green, blue, dark-blue, brown, 
dark-green, etc.
}
\label{spe_266Rf} 
\end{figure*}

In order to examine the competition between spontaneous fission and 
$\alpha$-decay, we have computed the $\alpha$-decay half-lives t$_{\alpha}$
using  the Viola-Seaborg formula \cite{Viola-Seaborg}
\begin{equation} \label{VSeaborg-new}
\log_{10} t_{\alpha} =  \frac{AZ+B}{\sqrt{ {\cal{Q}}_{\alpha}}} + CZ+D 
\end{equation}
with the parameters $A$, $B$, $C$ and  $D$  given in 
Ref.~\cite{TDong2005}. The ${\cal{Q}}_{\alpha}$ value is obtained in 
each case from the calculated binding energies. 

We have kept axial symmetry as a selfconsistent symmetry along the 
whole fission path of each of the studied nuclei to reduce the already 
substantial computational effort. We are aware of the role of 
triaxiality around the top of the inner fission barrier, but as 
discussed below it seems to have little impact on spontaneous fission 
half lives in the nuclei where this effect has been studied. Typically, once the $\gamma$ degree of freedom is taken 
into account, the inner barriers are a few (2-3) MeV smaller than the axial 
ones \cite{Rayner-fission-1,Delaroche-2006,Abusara-2010,Agbemava-Hyperheavy-2019}. Nevertheless, 
it has also been shown that the lowering of the inner barrier comes along with
a decrease of the level density, and therefore of the pairing correlations that
leads to an increase in the collective inertia 
\cite{Baran-1981,Bender-1998,Delaroche-2006} that tends to compensate  the effect of the lower barrier in the final 
value of the action. As a consequence, 
the impact of triaxiality on the spontaneous fission half-lives is very 
limited in the nuclei considered, mostly actinides and $N < 190$. Although a careful analysis is still required, we expect that the same
trend will be valid for all the nuclei considered in this work. 
Moreover, in the framework of the least 
action approach, it has been shown that pairing fluctuations  play a 
key role to improve the comparison between the theoretical and 
experimental t$_\mathrm{SF}$ values 
\cite{Min_Action_Gogny,Min_Action_GognyRayner2018,Min_Action_RMF} and, 
at the same time, they can restore axial symmetry along the fission 
path \cite{Sad14,Min_Action_Skyrme_2}. 

Concerning the D1M$^{*}$ parametrization there has been some debate in the
literature \cite{spurious} about the spurious finite-size instabilities that appear when
this parametrization is used in HFB solvers in coordinate space. As discussed at
length in \cite{spurious2} we have never observed such a behavior in our
calculations using the harmonic oscillator basis. The conclusion of \cite{spurious2,spurious3}
is that the ultraviolet cutoff characteristic of the harmonic oscillator basis regularizes 
the spuriosities of the parametrization. Nevertheless, we have carried out
some convergence tests with the basis size for the binding energy and found
that the energy gains obtained with D1M$^{*}$ are very close to the ones of
D1M that is supposed to be free from spuriosities \cite{spurious}. The similitude
between the D1M and D1M$^{*}$ results obtained in this paper for nuclei close
to the stability line (see, Sec.~\ref{RESULTS}) allow us to conclude that our calculations are free 
from spurious effects.

%

\begin{figure*}
\includegraphics[width=0.95\textwidth]{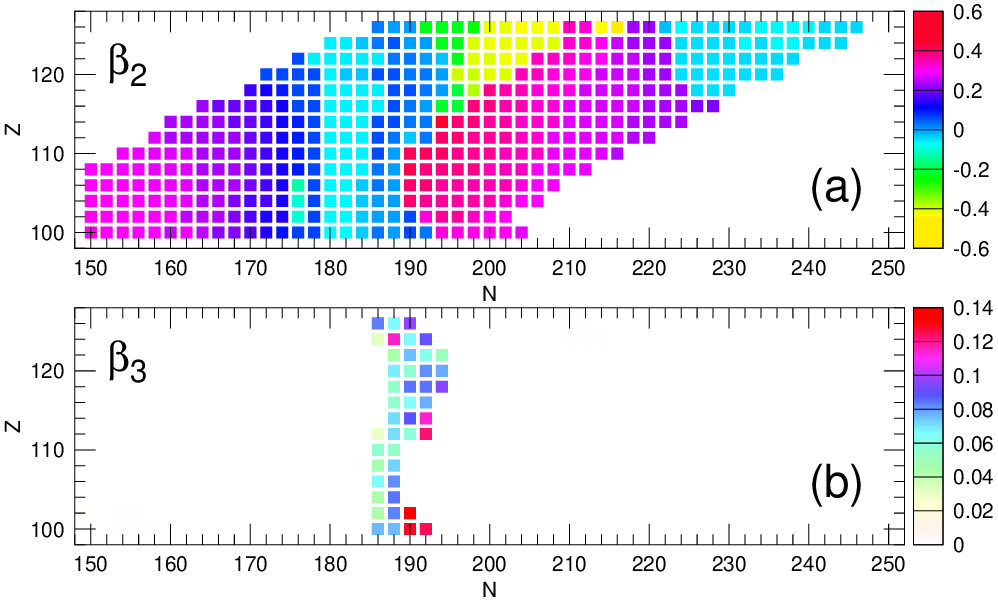} \\
\includegraphics[width=0.95\textwidth]{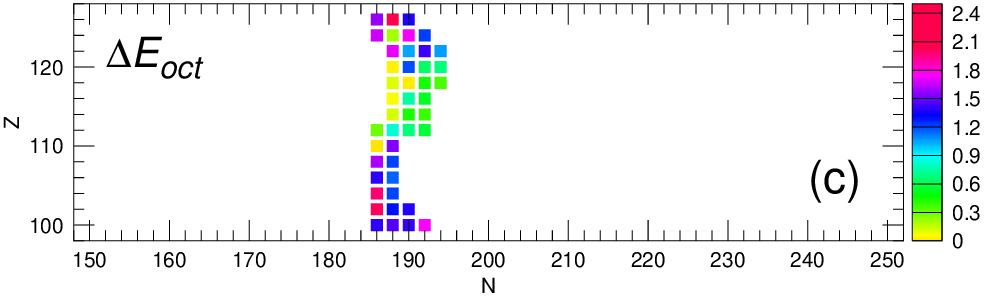}  
\caption{ 
(Color online) Deformation parameters $\beta_{2}$ [panel (a)] and 
$\beta_{3}$ [panel (b)] for the ground states of the even-even 
superheavy nuclei studied in this work.
The octupole correlation
energies $\Delta$ E$_{oct}$ corresponding to the nuclei shown 
in panel (b) are plotted in panel (c).
 Results have been obtained with 
the Gogny-D1M$^{*}$ EDF. For more details, see the main text.
}
\label{map} 
\end{figure*} 

%
%
%

\section{Discussion of the results}
\label{RESULTS}
In this section, we present  the results of our fission calculations 
with Gogny-D1M$^{*}$. In Sec.~\ref{validate}, we discuss the D1M$^{*}$ 
results for a set of U, Pu, Cm, Cf, Fm, No, Rf, Sg and Fl nuclei for 
which experimental data are available 
\cite{Refs-barriers-other-nuclei-1,Refs-barriers-other-nuclei-2,Refs-barriers-other-nuclei-3-tsf,Bertsch15}. 
In Sec.~\ref{methodology}, we illustrate the methodology employed to 
compute fission observables for the nuclei $^{256}$No and $^{298}$No. 
The same methodology has been used for all the nuclei studied in this 
work. The aim of Secs.~\ref{validate} and \ref{methodology} is to 
validate the Gogny-D1M$^{*}$ EDF as a reasonable choice for fission 
studies. We will also compare our results with those obtained with the 
Gogny-D1M EDF. In Sec.~\ref{methodology}, we will briefly  discuss the 
neutron and proton single-particle energies, as functions of the 
quadrupole moment Q$_{20}$,  in the case of $^{266}$Rf which is taken 
as an illustrative example. The systematic, provided by the 
Gogny-D1M$^{*}$ HFB approach, for the fission paths, t$_\mathrm{SF}$ 
and t$_{\alpha}$ values is presented in Sec. \ref{FB-systematcis}. In 
addition, ground state deformations, pairing correlations, two-nucleon 
separation energies and barrier heights are discussed in this section.

\subsection{Heavy nuclei with known  experimental data}
\label{validate}

In this section, we discuss the results obtained with the 
Gogny-D1M$^{*}$ EDF for the set of nuclei $^{232-238}$U, 
$^{238-244}$Pu, $^{240-248}$Cm, $^{250,252}$Cf, $^{250-256}$Fm, 
$^{252-256}$No, $^{256-260}$Rf, $^{258-262}$Sg, $^{264}$Hs and 
$^{286}$Fl for which, experimental data are available 
\cite{Refs-barriers-other-nuclei-1,Refs-barriers-other-nuclei-2,Refs-barriers-other-nuclei-3-tsf,Bertsch15}. 
Calculations for those nuclei have been carried out along the lines 
discussed in 
Refs.~\cite{Rayner-fission-1,Rayner-fission-2,Rayner-fission-3} (see, 
also Sec.~\ref{methodology}). Previous theoretical results, based on 
the parameter sets D1S and D1M, can be found in 
Refs.~\cite{Delaroche-2006,WERP02,Warda-Egido-2012,Rayner-fission-1}.

In Table~\ref{Table1}, we compare the predicted heights 
$B_{I}^{th}$(D1M$^{*}$) and $B_{II}^{th}$(D1M$^{*}$) for the inner and 
outer  barriers as well as the excitation energies 
$E_{II}^{th}$(D1M$^{*}$) of the fission isomers with the experimental 
ones $B_{I}^{exp}$, $B_{II}^{exp}$ and $E_{II}^{exp}$ 
\cite{Refs-barriers-other-nuclei-1,Refs-barriers-other-nuclei-2}. 
Results obtained with the Gogny-D1M EDF \cite{Rayner-fission-1} are 
also included in the table for comparison. All the theoretical values 
have been obtained from the energies $E_{HFB} + E_{ROT}$ by looking at 
the energy differences between the ground state and the corresponding 
configurations along the fission paths of the studied nuclei. The 
calculated heights $B_{I}^{th}$(D1M$^{*}$) and $B_{I}^{th}$(D1M) are 
always larger than the experimental ones. The maximal deviation 
$B_{I}^{th}$(D1M) - $B_{I}^{exp}$ = 4.88 MeV occurs for $^{248}$Cm 
while the largest difference $B_{I}^{th}$(D1M$^{*}$) - $B_{I}^{exp}$ = 
4.87 MeV is obtained for $^{246}$Cm. When triaxial shapes are allowed 
at and in the vecinity of the first barrier we get a reduction of the
barrier height. The reduction (see values in second row for the triaxial
barrier heights) depends on the nucleus and goes from a few hundred keV up to 
$\sim$ 3 MeV. However, the inclusion of triaxial effects cannot 
reconcile the theoretical values with the experimental ones. Given the 
reasonable agreement with the experimental t$_\mathrm{SF}$ values 
discussed below we can conclude that there is an unsettled issue with 
the comparison between the fission barrier heights obtained in the 
least energy framework and the experimental values (see 
\cite{Min_Action_GognyRayner2018} for a discussion in the context of 
the least action path method). It has to be considered that the 
experimental values are obtained in a model dependent manner from 
induced fission cross sections. The comparison of fission barrier 
heights, along with the role played by the width of the barrier and the 
collective inertias in the observables, undoubtedly deserve further 
consideration.

The discrepancies can be 
attributed both to the use of an axially symmetric path and also to the 
uncertainties in the model dependent methodology used to extract the 
experimental values. It is also worth to point out that neither of the 
two Gogny forces were specifically fitted to fission barrier data. 
Nevertheless, the global trend of the computed quantities agrees 
reasonably well with experimental data and the results of other 
calculations 
\cite{UNEDF1,Abu-2012-bheights,Robledo-Giulliani,Delaroche-2006}.
 
%
\begin{figure*}
\includegraphics[width=0.33\textwidth]{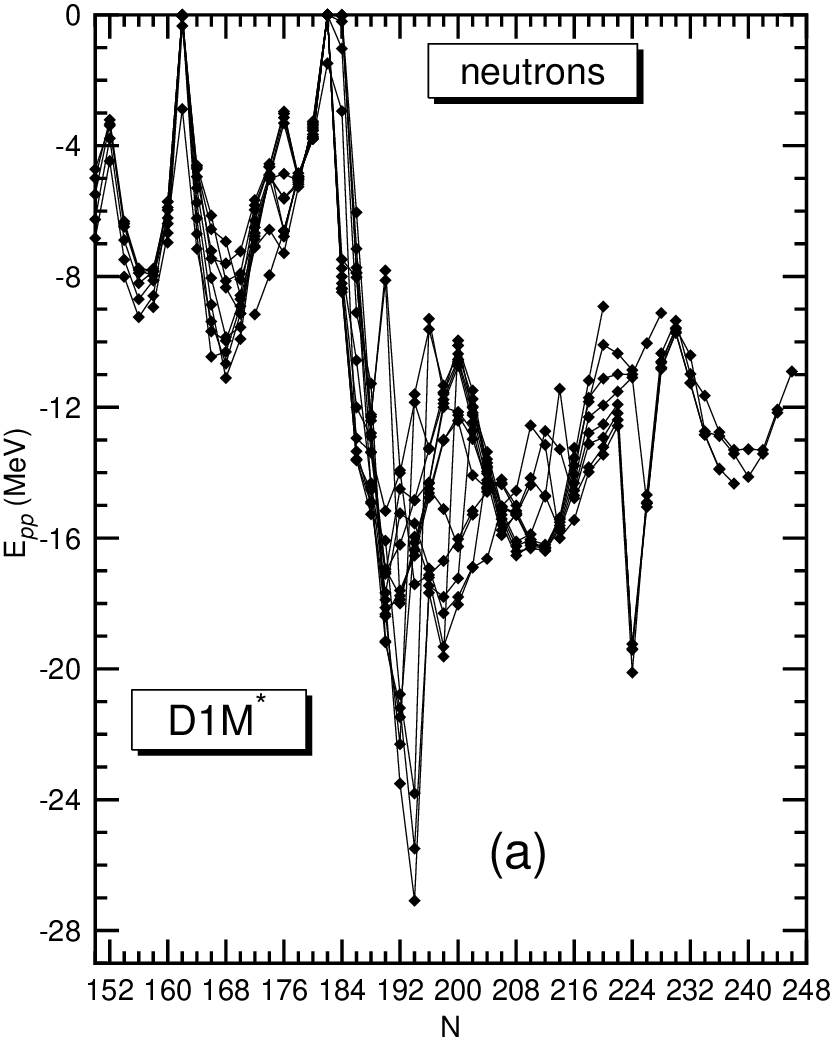} 
\includegraphics[width=0.32\textwidth]{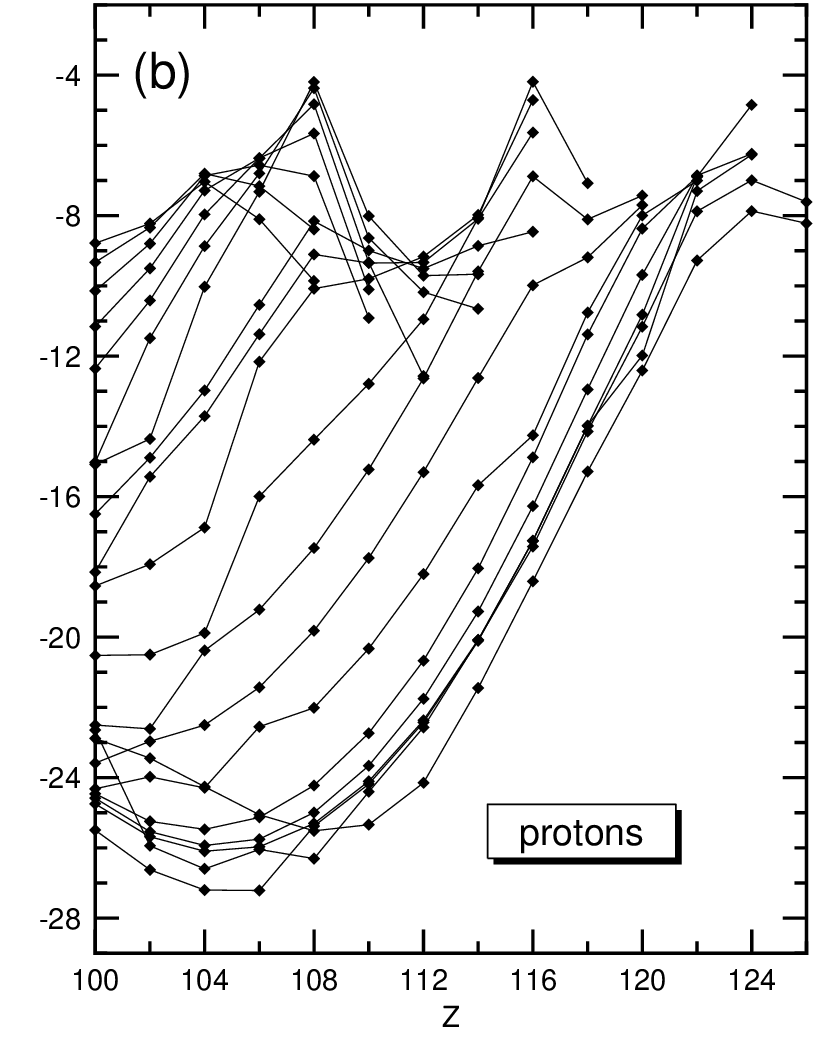}
\includegraphics[width=0.32\textwidth]{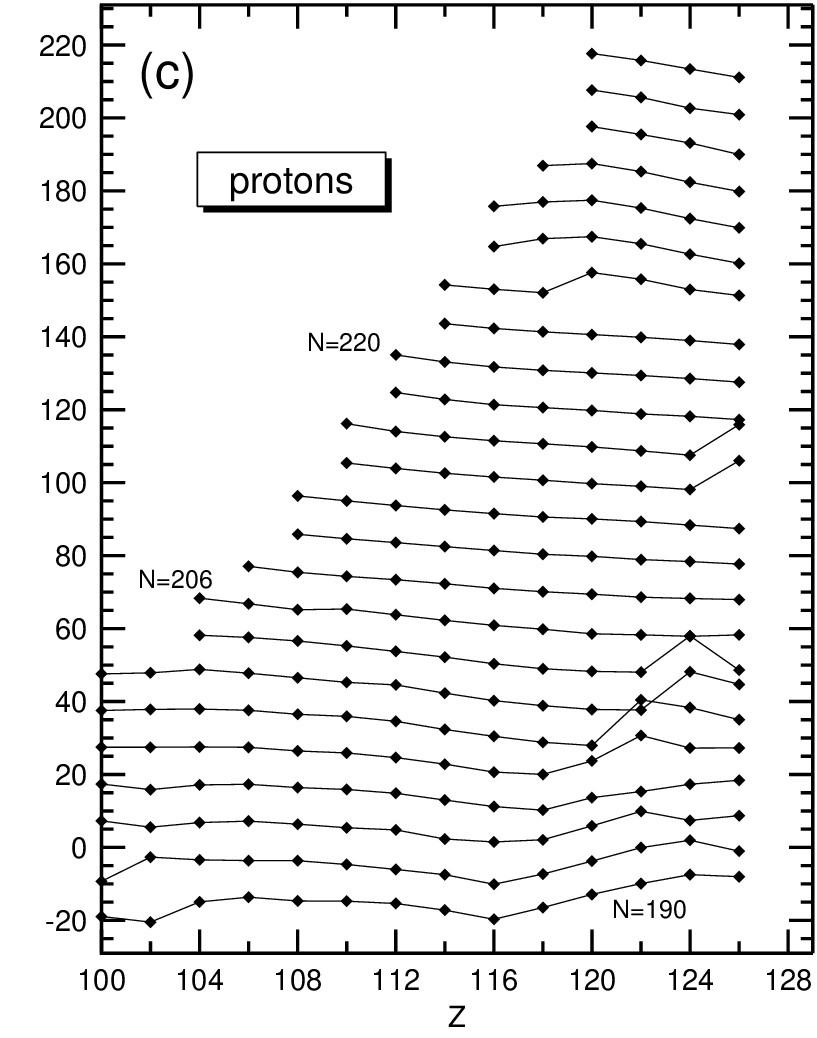}
\caption{The neutron  and proton 
pairing interaction energies corresponding to the ground states 
of the studied nuclei are plotted 
in panel (a) [panels (b) and (c)]. Starting with the N =192 isotones
in panel (c), the curves have been shifted by 10 MeV. Results have been obtained
with the Gogny-D1M$^{*}$ EDF. For more details, see the main text. 
}
\label{pair-FIG} 
\end{figure*}  

For the outer barrier heights and the excitation energies of the 
fission isomers, we obtain similar results in the two calculations, 
although the values obtained with D1M tend to be slightly larger than 
those obtained with D1M$^{*}$. This trend is a consequence of the more 
gentle decline observed in the Gogny-D1M fission paths for quadrupole 
moments Q$_{20}$ $\ge$ 40 b (see, also Sec.~\ref{methodology}). 
Reflection symmetry is allowed to break along the fission path whenever 
it is energetically favorable. For the nuclei studied in this section, 
octupole correlations reduce the outer barrier heights by a few MeV. 
However, in the case of $^{244}$Pu, we still observe deviations of up to $B_{II}^{th}$(D1M) - 
$B_{II}^{exp}$ = 4.75 MeV and $B_{II}^{th}$(D1M$^{*}$) - $B_{II}^{exp}$ 
= 4.18 MeV, respectively. The previous 
results, as well as the ones obtained with the D1S pamaterization 
\cite{Delaroche-2006}, seem to suggest that effects other than the ones 
associated with shape degrees of freedom might be required to improve 
the comparison with the $B_{II}^{exp}$ values. However, we would like 
to stress again the model dependent character of the $B_{II}^{exp}$ 
values, making these quantities less reliable for a comparison with 
theoretical predictions, than real observables like fission half-lives 
. In the case of the fission isomers, we observe for $^{242}$Pu differences of up to  
$E_{II}^{th}$(D1M) - $E_{II}^{exp}$ = 1.37 MeV and 
$E_{II}^{th}$(D1M$^{*}$) - $E_{II}^{exp}$ = 0.70 MeV, respectively.

In Fig.~\ref{tsf-nuclei-with-data}, the spontaneous fission half-lives 
t$_\mathrm{SF}$ obtained for the nuclei $^{232-238}$U, $^{240}$Pu, 
$^{248}$Cm, $^{250}$Cf, $^{250-256}$Fm, $^{252-256}$No, $^{256-260}$Rf, 
$^{258-262}$Sg, $^{264}$Hs and $^{286}$Fl are plotted as functions of 
the fissility-related parameter $Z^{2}$/A. Results obtained within both the GCM 
and ATDHFB schemes are shown for both the Gogny-D1M$^{*}$  [panel (a)] 
and the Gogny-D1M [panel (b)] EDFs. The results are compared with the 
corresponding experimental data 
\cite{Refs-barriers-other-nuclei-3-tsf,Bertsch15}. Calculations have been carried 
out with E$_{0}$ = 0.5 MeV and E$_{0}$ =1.0 MeV. This is a very 
delicate comparison as the experimental t$_\mathrm{SF}$ values span 27 
orders of magnitude. Regardless of the considered Gogny-EDF, the 
computed half-lives display an even larger variability depending on the 
scheme  employed. As it has already been noted in previous studies 
\cite{Rayner-fission-1,Rayner-fission-2,Rayner-fission-3}, increasing 
the value of E$_{0}$ improves the comparison with the experiment in 
both of the two schemes especially for Fm, No, Rf, Sg, Hs and Fl 
nuclei. It is satisfying to observe that D1M$^{*}$  provides the same 
quality of results as D1M and  captures the pronounced experimental 
reduction in the spontaneous fission half-lives for increasing values 
of $Z^{2}$/A. Moreover, along isotopic chains the trend with neutron 
number is also reasonably well reproduced by both EDFs.

\subsection{The nuclei $^{256}$No and $^{298}$No}
\label{methodology}

In this section, we illustrate the methodology employed to compute the 
fission paths and other fission-related quantities for the nuclei 
$^{256}$No and $^{298}$No. The corresponding Gogny-D1M$^{*}$ HFB plus 
the zero point rotational energies are plotted in panels (a) and (e) of 
Fig.~\ref{FissionBarriers_pedagogical} as functions of the quadrupole 
moment Q$_{20}$. Results obtained with the Gogny-D1M EDF are also 
included in each of the panels. The zero point vibrational energies are 
not included in the plots as they are rather constant as functions of 
the quadrupole moment. However, we always consider such vibrational 
corrections in the computation of the t$_\mathrm{SF}$ and t$_{\alpha}$ 
lifetimes.

In the case of $^{256}$No, the absolute minimum is located at
Q$_{20}$ = 16b. The fission isomer, located at 
Q$_{20}$ = 52b, lies 
$E_{II}^{th}$(D1M$^{*}$)= 1.16 MeV above the ground 
state from which it is separated by the inner barrier 
the top of which, is located at 
Q$_{20}$ = 36b with the height 
$B_{I}^{th}$(D1M$^{*}$)= 9.53 MeV. The outer barrier 
with the height 
$B_{II}^{th}$(D1M$^{*}$)= 3.20 MeV 
is found at 
Q$_{20}$ = 72b. For the
Gogny-D1M fission path the absolute minimum is also located  at
Q$_{20}$ = 16b
while 
$B_{I}^{th}$(D1M)= 9.46 MeV  (Q$_{20}$ = 36b), 
$E_{II}^{th}$(D1M)= 1.51 MeV (Q$_{20}$ = 52b) and 
$B_{II}^{th}$(D1M)= 3.90 MeV (Q$_{20}$ = 72b).  

%
\begin{figure*}
\includegraphics[width=0.47\textwidth]{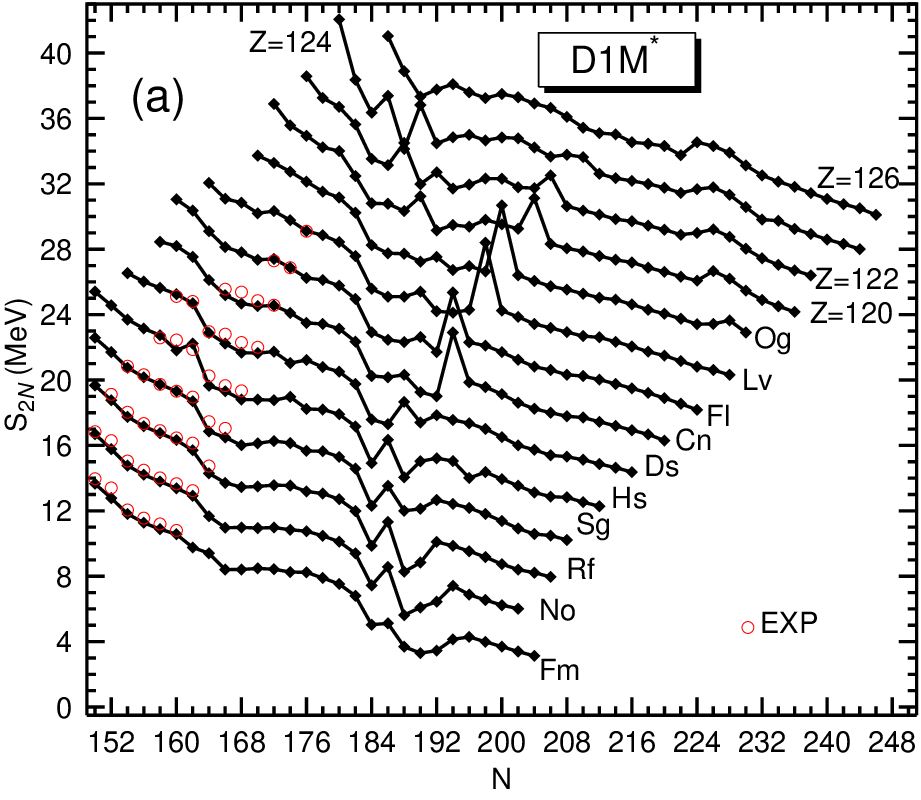} 
\includegraphics[width=0.48\textwidth]{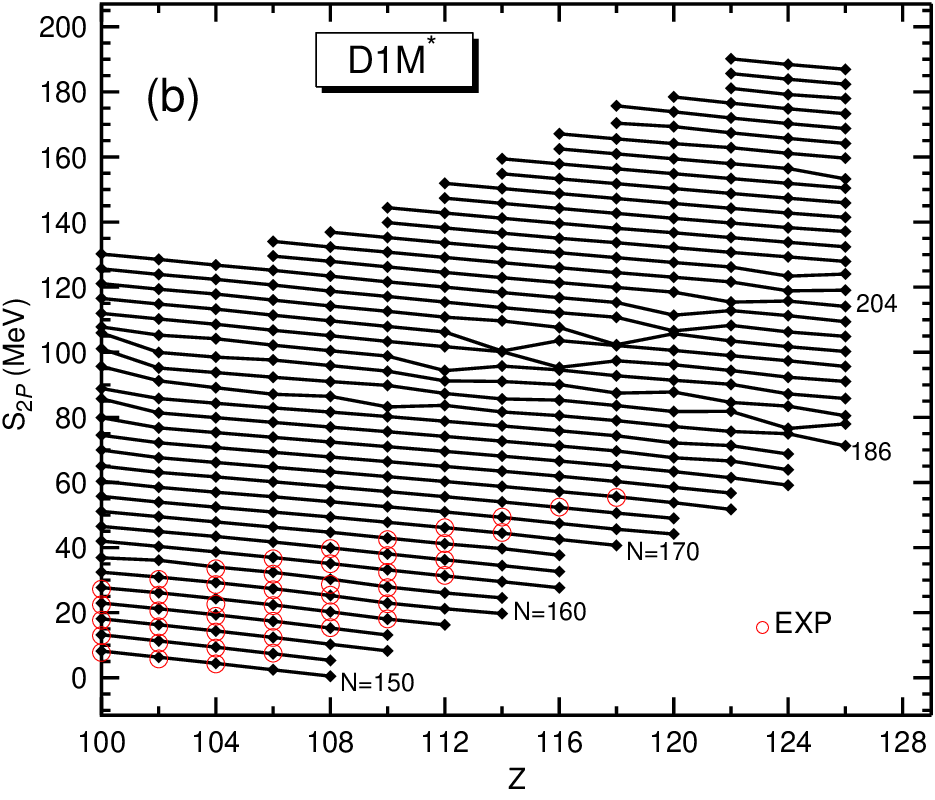}
\caption{(Color online) The two-neutron  (two-proton) separation 
energies are plotted as functions of the neutron number N (proton 
number Z) in panel (a) [panel (b)]. Starting with the Z = 102 isotopes 
in panel (a) [N = 152 isotones in panel (b)] the curves have been 
shifted by 2 MeV (4 MeV). Results have been obtained with the 
Gogny-D1M$^{*}$ EDF. Experimental data have been taken from 
Ref.~\cite{S2N-S2P-EXP}. For more details, see the main text.
}
\label{two-nucleon-sep-ener-FIG} 
\end{figure*}

For the neutron-rich nucleus $^{298}$No, the absolute minimum is 
located at Q$_{20}$ = 24b. We have obtained $B_{I}^{th}$(D1M$^{*}$)= 
4.00 MeV  (Q$_{20}$ = 42b), $E_{II}^{th}$(D1M$^{*}$)= 1.21 MeV  
(Q$_{20}$ = 52b) and $B_{II}^{th}$(D1M$^{*}$)= 2.53  MeV  (Q$_{20}$ = 
60b). On the other hand, for the Gogny-D1M fission path the absolute 
minimum is also located at Q$_{20}$ = 24b while $B_{I}^{th}$(D1M)= 4.20 
MeV  (Q$_{20}$ = 42b), $E_{II}^{th}$(D1M)= 1.85 MeV  (Q$_{20}$ = 52b) 
and $B_{II}^{th}$(D1M)= 3.32 MeV  (Q$_{20}$ = 60b).

From the previous results one realizes that, for both $^{256}$No and 
$^{298}$No, the topology of the fission paths obtained with the D1M 
and D1M$^{*}$ Gogny-EDFs are quite similar albeit with a more gentle 
decline of the former as compared with the latter for  quadrupole 
moments Q$_{20}$ $\ge$ 40b [see, for example, panel (a)]. However, in 
the case of $^{298}$No, there is a pronounced  under-binding for all the 
configurations along the Gogny-D1M$^{*}$ path. Such a shift with 
increasing neutron number N along an isotopic chain has been pointed 
out in Ref.~\cite{gogny-D1MSTAR} and it is a direct consequence of the 
different density dependence of the symmetry energy  in the two EDFs.

The octupole and hexadecupole  moments obtained for $^{256}$No and 
$^{298}$No are plotted in panels (b) and (f) of the figure. In the case 
of $^{256}$No, octupole deformed  shapes are energetically favored for 
-14b $\le$  Q$_{20}$ $\le$ 4 b and 24b $\le$  Q$_{20}$ $\le$ 34 b. 
Octupole correlations also play a prominent role for Q$_{20}$ $\ge$ 80b 
leading to a reflection-asymmetric outer sector of the fission path 
much lower in energy than the symmetric (i.e., Q$_{30}$ = 0) one. In 
the case of $^{298}$No, octupole deformed shapes play a role around 
both the spherical configuration and Q$_{20}$ = 60b while for larger 
quadrupole moments the (minimal energy) path corresponds to 
reflection-symmetric shapes.

The pairing interaction energies E$_{pp}= - 1/2 \mathrm{Tr} \left( \Delta 
\kappa^{*}\right)$ \cite{rs} are depicted in panels (c) and (g) of 
Fig.~\ref{FissionBarriers_pedagogical}. The proton and neutron pairing 
content of the two Gogny-EDFs is similar because the two combinations 
of parameters $W_{i}-B_{i}-H_{i}+M_{i}$ ($i=1,2$) of the finite-range 
part of the Gogny-EDF \cite{gogny-d1s} which govern the pairing 
strength have been kept fixed to their D1M values in the fitting 
protocol of D1M$^{*}$ \cite{gogny-D1MSTAR}. The collective masses are 
plotted in panels (d) and (h). Their evolution, as functions of 
Q$_{20}$, is well correlated with the one of the pairing energies shown 
in panels (c) and (g). Regardless of the considered EDF, the GCM and 
ATDHFB masses exhibit a similar pattern, though the values of the 
latter are always larger than the former. As a consequence, the action 
Eq. (\ref{Action}) computed within the ATDHFB scheme is larger than the 
GCM one. Those differences can represent a change of several orders of 
magnitude in the predicted fission half-lives Eq. (\ref{TSF}) 
\cite{Robledo-Giulliani,Rayner-fission-1,Rayner-fission-2,Rayner-fission-3,Rayner-fission-4,Rayner-fission-5,Min_Action_Gogny,Min_Action_GognyRayner2018}. 
This is the reason to consider both kinds of collective masses in this 
work. For example, in the case of $^{256}$No and E$_{0}$ = 0.5 MeV, we 
have obtained the values $\log_{10}$t$_\mathrm{SF}^{GCM}$ = 8.88
and  $\log_{10}$t$_\mathrm{SF}^{ATDHFB}$ = 12.04  with the 
Gogny-D1M$^{*}$ EDF. For the same nucleus, the values obtained with D1M 
are $\log_{10}$t$_\mathrm{SF}^{GCM}$ = 10.53  and  
$\log_{10}$t$_\mathrm{SF}^{ATDHFB}$ = 13.72. A larger E$_{0}$ value 
leads to a reduction in the predicted fission half-lives. For 
$^{256}$No and E$_{0}$ = 1.0 MeV, we have obtained 
$\log_{10}$t$_\mathrm{SF}^{GCM}$ = 6.66 and  
$\log_{10}$t$_\mathrm{SF}^{ATDHFB}$ = 9.02  with D1M$^{*}$ while the 
corresponding values with D1M are $\log_{10}$t$_\mathrm{SF}^{GCM}$ = 
8.18 and $\log_{10}$t$_\mathrm{SF}^{ATDHFB}$ = 10.67.

%
\begin{figure*}
\includegraphics[width=1.0\textwidth]{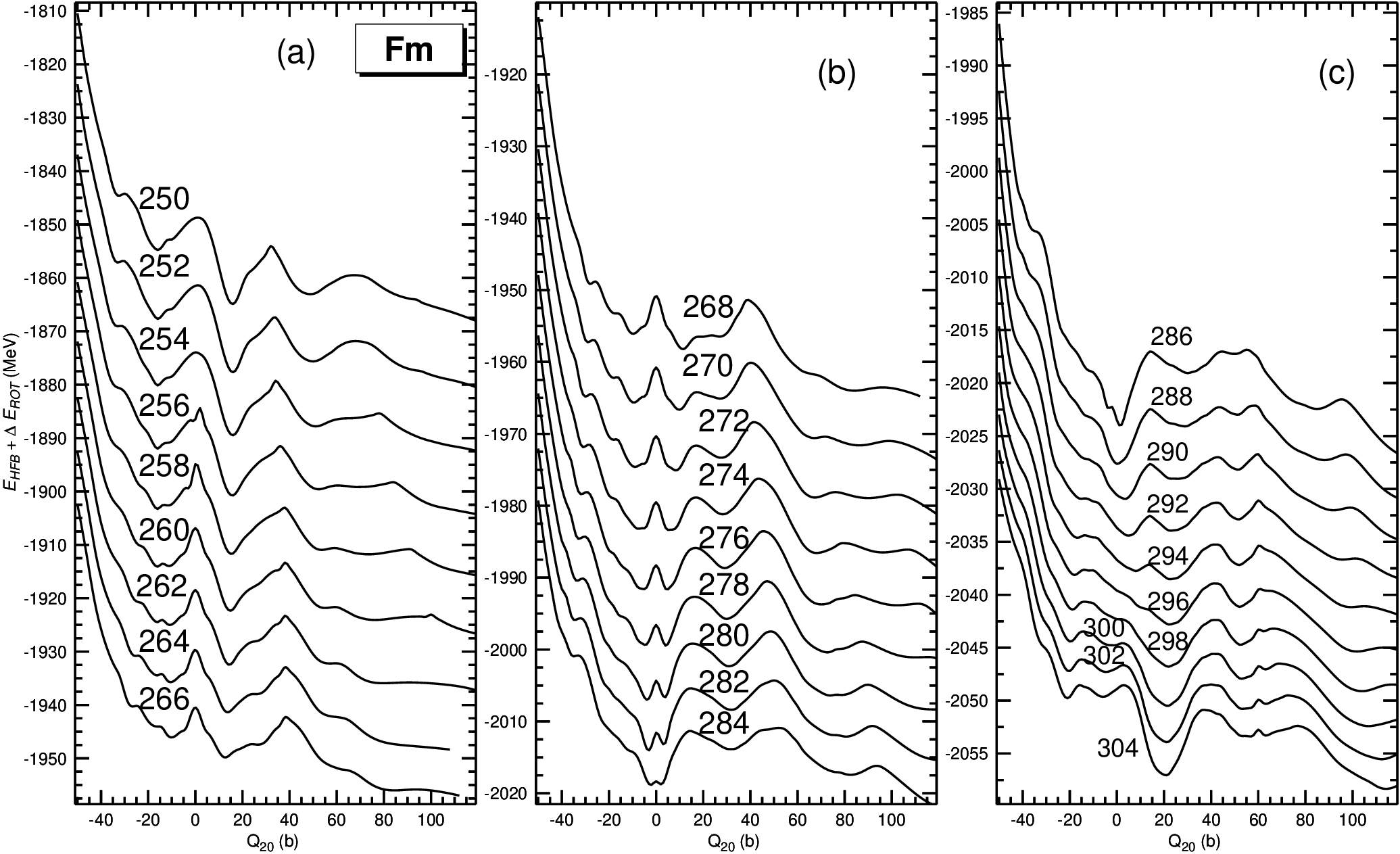} 
\caption{ 
The HFB plus the zero point rotational energies, obtained with the  
Gogny-D1M$^{*}$ EDF, are plotted  as functions of the quadrupole moment 
Q$_{20}$ for the nuclei $^{250-304}$Fm. Starting from $^{252}$Fm [panel 
(a)], $^{270}$Fm [panel (b)] and $^{288}$Fm [panel (c)], the curves 
have been successively  shifted by 2.5 MeV in order to accommodate them 
in a single plot. Note that the energy scales span different ranges in 
each panel. For more details, see the main text. 
}
\label{FissionBarriersFm} 
\end{figure*}

To summarize the results already discussed in Secs.~\ref{validate} and 
\ref{methodology}, it has been shown that, in spite of the  theoretical 
uncertainties in the models employed to describe the considered 
quantities, the Gogny-D1M$^{*}$ HFB framework describes reasonably well  
the trend with mass number in the studied observables and also that its 
predictions compare well with those obtained with the parametrization 
D1M. This validates the use of the D1M$^{*}$ parameter set  to study 
the systematic of the fission paths as well as other relevant 
quantities for the even-even nuclei shown in Fig.~\ref{studied_nuclei}.

Before concluding this section, we turn our attention to 
single-particle properties. As it is well known, the existence of 
minima as a function of some collective parameter is strongly linked
to the existence of 
regions with low single-particle densities (Jahn-Teller effect). 
Therefore, plotting single-particle energies (SPEs) as a function of the
quadrupole deformation helps us to 
identify regions where energy gaps favor the appearance of deformed 
minima. As in our mean-field calculations we solve the full HFB 
equations, the only quantities that can be properly defined are the 
quasiparticle energies \cite{rs}. However, in order to have the usual 
Nilsson-like type of diagram we have chosen to plot the eigenvalues of the 
Routhian $h = t + \Gamma - \lambda_{Q_{20}} Q_{20} - \lambda_{Q_{30}} 
Q_{30}$, with t being the kinetic energy and $\Gamma$ the Hartree-Fock 
field. The term $\lambda_{Q_{20}} Q_{20} + \lambda_{Q_{30}} Q_{30}$ 
contains the Lagrange multipliers used to enforce the corresponding 
quadrupole and octupole constrains. 
 
The neutron and proton SPEs, computed with the Gogny-D1M$^{*}$ EDF, are 
plotted in Fig.~\ref{spe_266Rf} for $^{266}$Rf. As can be seen from 
panel (a), a spherical sub-shell is predicted at N = 164. However, the 
spherical shell closures at N = 126 and N = 184 are much more 
pronounced. Note, that in our calculations the spherical magic number 
next to  N = 184 is N = 228. Prolate (oblate) gaps are observed at N = 
152 and N = 162 (N = 172). In the case of protons [panel (b)], we 
observe a  sub-shell at Z = 114  smaller than the shell closures at Z = 
92, 120 and 126. Prolate proton gaps are also found at Z = 104 and Z = 
108. Similar results are obtained with the Gogny-D1M EDF and will not 
be shown here. The previous results also agree well with those obtained 
with the Gogny-D1S 
parametrization \cite{Warda-Egido-2012}. For a discussion of 
spherical and deformed shell gaps 
in superheavy nuclei
within the 
relativistic mean-field (RMF) 
and Skyrme-EDF approximations the reader is
referred, for example, to Refs.~\cite{RMF-coco1,Bender-SPE}.

\subsection{Systematic of the fission paths and spontaneous fission half-lives}
\label{FB-systematcis}

For each of the even-even nuclei shown in Fig.~\ref{studied_nuclei}, we 
have performed a systematic analysis of the fission path, obtained with 
the Gogny-D1M$^{*}$ EDF, along the lines discussed in 
Sec.~\ref{methodology}. Let us first turn our attention to the ground 
state quadrupole $\beta_{2}$ Eq.(\ref{beta2def}) and octupole 
$\beta_{3}$ Eq. (\ref{beta3def})  deformation parameters. They  are 
shown in panels (a) and (b) of Fig.~\ref{map}. A more detailed 
description of the fission paths will be presented later on in the 
paper.

As can be seen from panel (a) of the figure, well deformed prolate 
ground states (0.05 $\le$ $\beta_{2}$ $\le$ 0.29) are obtained for 
nuclei with  proton and neutron numbers 100 $\le$ Z $\le$ 122 and 150 
$\le$ N $\le$ 178. There are a few exceptions in the region 
corresponding to the well deformed oblate nuclei $^{278}$No, 
$^{280}$Rf, $^{282}$Sg and $^{300}$122. A region of weakly deformed 
and/or spherical ground states (-0.06 $\le$ $\beta_{2}$ $\le$ 0.06) 
emerges as we approach  N=184 and roughly extends up to N = 192-194. 
Beyond N=192-194, our calculations predict a region with pronounced 
prolate (0.21 $\le$ $\beta_{2}$ $\le$ 0.37) deformations. A small 
pocket of strongly oblate (-0.41 $\le$ $\beta_{2}$ $\le$ -0.35) systems 
is also found in this region. Finally, weakly oblate ($\beta_{2}$ 
$\approx$ -0.04)  ground states are found for 118 $\le$ Z $\le$ 126 and 
224 $\le$ N $\le$ 246. In our calculations, specially with increasing Z 
values, the fission paths obtained for some neutron-rich nuclei exhibit 
a very complex topology with several competing minima. Similar results 
have been obtained in previous works 
\cite{Giuliani-Pinedo-SHE-rp,Warda-Egido-2012,cea_compilation}. 
Exception made of a small number of cases, in those situations where 
the identification of the absolute minimum of the fission path becomes 
more involved, we have taken the deepest minimum closest to sphericity 
as the ground state of the system \cite{cea_compilation}.

%
\begin{figure*}
\includegraphics[width=1.0\textwidth]{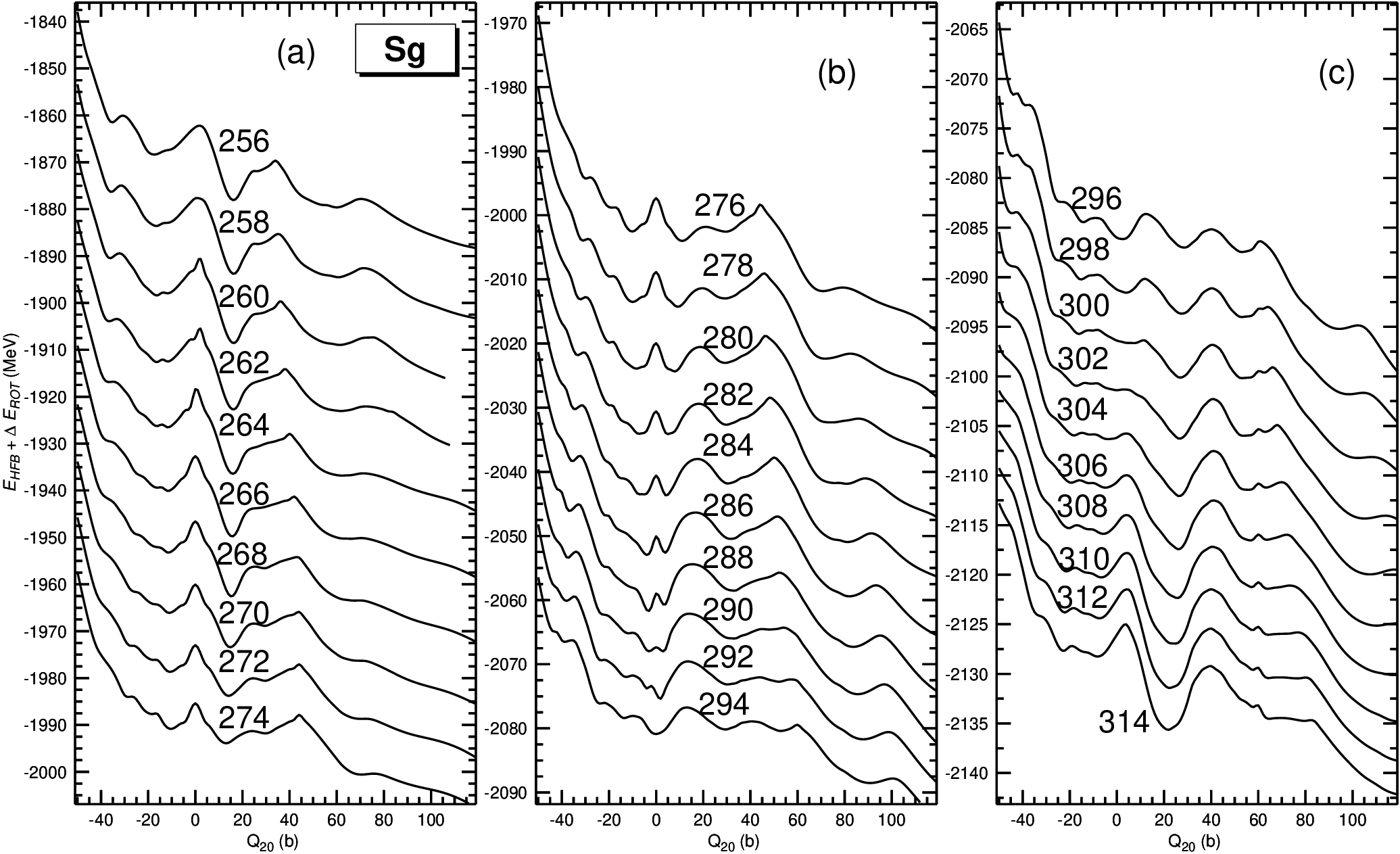} 
\caption{ 
The HFB plus the zero point rotational energies, obtained with the  
Gogny-D1M$^{*}$ EDF, are plotted  as functions of the quadrupole moment 
Q$_{20}$ for the nuclei $^{256-314}$Sg. Starting from $^{258}$Sg [panel 
(a)], $^{278}$Sg [panel (b)] and $^{298}$Sg [panel (c)], the curves 
have been successively  shifted by 2.5 MeV in order to accommodate them 
in a single plot. Note that the energy scales span different ranges in 
each panel. For more details, see the main text. 
}
\label{FissionBarriersSg} 
\end{figure*}

In most of the cases the ground state is reflection-symmetric, as can 
be seen in panel (b) of Fig.~\ref{map}. However, octupole deformed 
ground states are found for a small set of  nuclei with neutron numbers 
186 $\le$ N $\le$ 194. On the one hand, this agrees well, with previous results 
obtained 
with the Gogny-D1S \cite{Warda-Egido-2012} and BCPM 
\cite{Giuliani-Pinedo-SHE-rp} EDFs as well as within the Mac-Mic model 
\cite{At.Data-Moeller}. In particular, all these 
models predict the 
existence of an island of octupole deformed nuclei around N $\approx$ 196.   
Previous calculations based on the Skyrme-SLy6 EDF \cite{Erler2012} and four different 
RMF models \cite{Agbemava-2017} also agree on the existence of such islands of octupolarity 
in neutron-rich low-Z superheavy nuclei. However, the
 Skyrme-SLy6
and RMF calculations do not predict octupole deformed ground states 
for nuclei with atomic number Z $\ge$ 110. We have computed 
the octupole correlation energy \cite{Agbemava-2017}

\begin{equation}
\Delta E_{oct} = \Big| E(Q_{20},Q_{30}) - E(Q^{'}_{20},Q^{'}_{30}=0)\Big|
\end{equation} 
where E($Q_{20}$,$Q_{30}$) and E($Q^{'}_{20}$,$Q^{'}_{30}=0$) represent the 
binding energies of a given nucleus at the octupole deformed absolute 
minimum of the fission path and the absolute minimum obtained in 
reflection-symmetric calculations, respectively. The values of $\Delta$ E$_{oct}$ 
are plotted in panel (c) of Fig.~\ref{map}.
As can be seen, in our Gogny-D1M$^{*}$ calculations the largest 
octupole correlation energies are around 2 MeV. For example, we have obtained
$\Delta$ E$_{oct}$ = 2.02, 1.98 and 2.04 MeV for the nuclei
$^{288}$No, $^{290}$Rf and $^{314}$126.
 
The average values of higher multipolarity 
moments (for example, $\hat{Q}_{40}$ and $\hat{Q}_{60}$) are determined 
by the selfconsistent minimization procedure of the HFB energy. We have 
found, that the global trends observed in the corresponding $\beta_{4}$ 
and $\beta_{6}$ parameters compare  well with the ones obtained in 
previous  studies 
\cite{Warda-Egido-2012,Giuliani-Pinedo-SHE-rp,At.Data-Moeller,Muntian}.

The neutron pairing interaction energies corresponding to the ground 
states of the studied nuclei are plotted in panel (a) of 
Fig.~\ref{pair-FIG}. They vanish around  N = 162 and 184, which 
correlates well with the low densities of neutron SPEs in panel (a) of 
Fig.~\ref{spe_266Rf}. The smaller pairing correlations at N =162 are 
consistent with the prolate deformed ground states  obtained for the 
isotones $^{262}$Fm, $^{264}$No, $^{266}$Rf, $^{268}$Sg, $^{270}$Hs, 
$^{272}$Ds, $^{274}$Cn and $^{276}$Fl [see, panel (a) of 
Fig.~\ref{map}]. On the other hand, the large pairing energies around N 
= 156, 168, 194, 210, 224 and 240 are associated with regions of high 
density of SPE levels. 

%
\begin{figure*}
\includegraphics[width=1.0\textwidth]{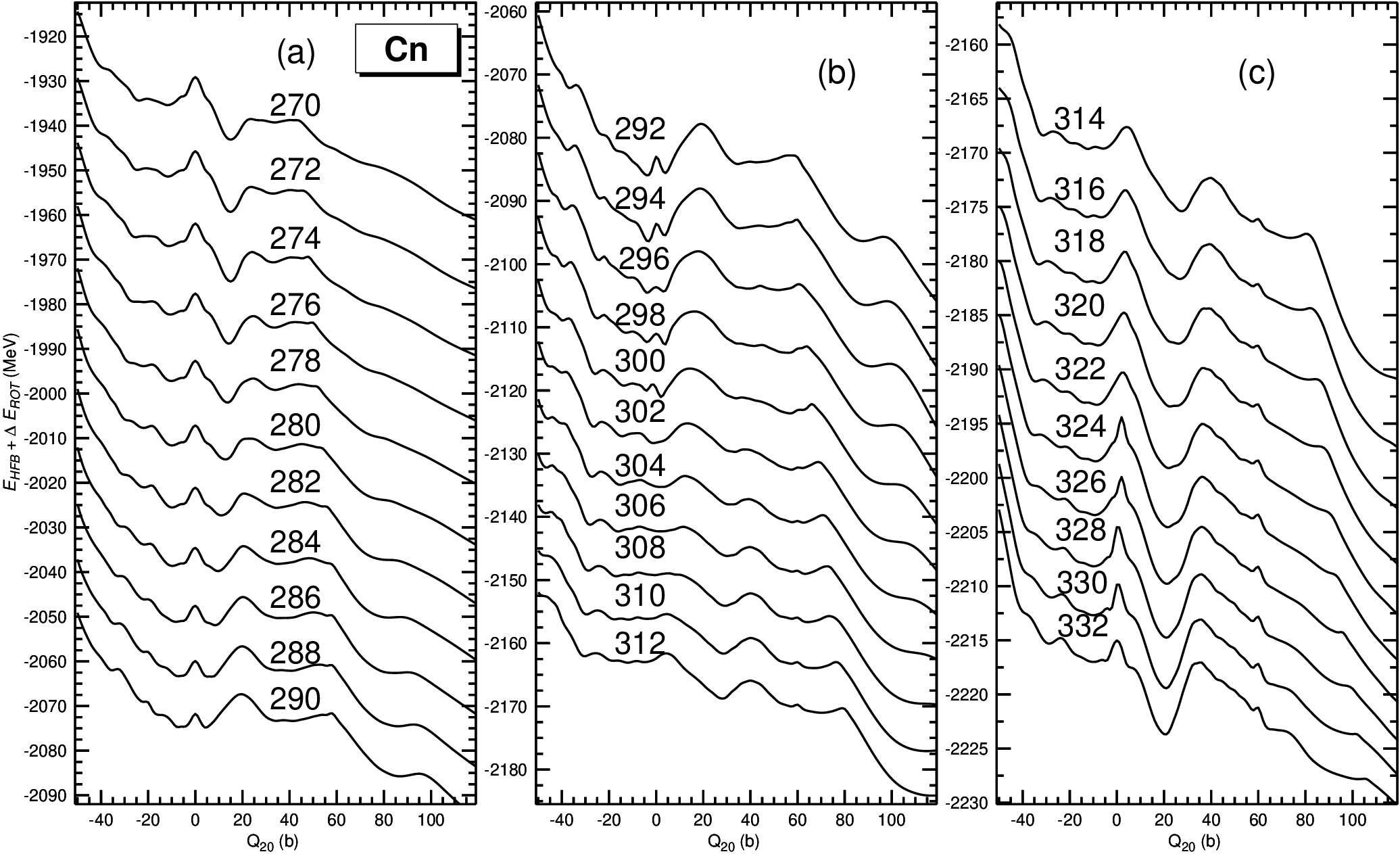} 
\caption{ 
The HFB plus the zero point rotational energies, obtained with the  
Gogny-D1M$^{*}$ EDF, are plotted  as functions of the quadrupole moment 
Q$_{20}$ for the nuclei $^{270-332}$Cn. Starting from $^{272}$Cn [panel 
(a)], $^{294}$Cn [panel (b)] and $^{316}$Cn [panel (c)], the curves 
have been successively  shifted by 2.5 MeV in order to accommodate them 
in a single plot. Note that the energy scales span different ranges in 
each panel. For more details, see the main text. 
}
\label{FissionBarriersCn} 
\end{figure*}

The pattern displayed by the proton pairing energies, shown in panels 
(b) and (c) of Fig.~\ref{pair-FIG}, is more involved than the one of 
the neutron pairing energies due to the fact that, for a given Z value 
and depending on neutron number, the isotopes can be  prolate, oblate 
and/or nearly spherical. 
The smaller pairing energies obtained for Z = 104, 108, 116, 120 and 
126  [panel (b)] are related to regions with low proton level 
densities. On the other hand, beyond N = 188 the proton pairing 
energies exhibit a change in tendency which can be associated to sudden 
ground state shape transitions. The proton pairing energies for N $\ge$ 
190 are depicted in panel (c). In order to avoid line crossings, 
starting with the N =192 isotones, the curves have been shifted by 10 
MeV. In the case of the N = 200 isotones, for example, the proton 
pairing energies decrease for 100 $\le$ Z $\le$ 120 and increase at Z = 
122. Such a behavior results from a prolate-to-oblate shape transition 
[see, panel (a) of Fig.~\ref{map}].

The two-neutron separation energies S$_{2N}$ are depicted in panel (a) 
of Fig.~\ref{two-nucleon-sep-ener-FIG} as functions of the neutron 
number N. Starting with the No isotopes (Z =102), the curves have been 
shifted by 2 MeV. With this choice of the energy shift the crossing of 
some lines is avoided and the general pattern of the S$_{2N}$ values is 
better  revealed. As expected, the two-neutron separation energies 
decrease with increasing neutron number since we move towards the 
corresponding two-neutron driplines. For each of the considered 
isotopic chains, we have  extended our calculations to very 
neutron-rich isotopes with S$_{2N}$ as low as  $\approx$ 4 MeV. For 
example, for the neutron-rich nuclei $^{304}$Fm, $^{314}$Sg, 
$^{332}$Cn,  $^{360}$122 and $^{372}$126 we have obtained 
S$_{2N}$ = 3.13, 4.22, 4.30, 4.40 and 4.10 MeV, respectively. As 
can be seen from the figure, there is a  sudden decline in the S$_{2N}$ 
values  around N = 162 and N=184. On the other hand prominent peaks (up 
to $\approx$ 4 MeV higher than their neighbors' values) have been 
obtained in the case of Cn, Fl, Lv, Og and Z=120 isotopes at N = 194, 
194, 198, 200 and 204, respectively. Those peaks reflect, the 
sensitivity of the S$_{2N}$ values with respect to the ground state 
shapes of the nuclei involved in their computation. In particular, for 
the already mentioned neutron numbers the peaks in the two-neutron 
separation energies are associated with sudden ground state shape 
changes [see, panel (a) of Fig.~\ref{map}]. A more realistic treatment 
of those peaks as well as the jumps when crossing shell closures 
requires the inclusion of dynamical correlations (i.e., 
symmetry-projected quadrupole configuration mixing 
\cite{Rayner-Sharma,global-q2-corre}) which are out of the scope of the 
present study. It is satisfying to observe that the Gogny-D1M$^{*}$ 
calculations reproduce reasonably well the available experimental data 
\cite{S2N-S2P-EXP}. Moreover, the Gogny-D1M$^{*}$ S$_{2N}$ values 
compare well with the ones obtained with the D1M parametrization. For 
example, for the isotopes $^{250-256}$Fm, we have obtained S$_{2N}$ = 
13.65, 12.77, 11.81, and 11.26 MeV with D1M$^{*}$ and S$_{2N}$ = 14.00, 
13.14, 12.23 and 11.73 MeV with D1M. All these theoretical values 
should be compared with the experimental ones \cite{S2N-S2P-EXP} i.e., 
S$_{2N}$ = 13.97, 13.40, 12.06 and 11.56 MeV. On the other hand, for the 
very neutron-rich isotopes $^{302,304}$Fm we have obtained 
S$_{2N}$ = 3.39 and 3.12 MeV with D1M$^{*}$ and 
S$_{2N}$ = 4.00 and 3.70 MeV with D1M.

%
\begin{figure*}
\includegraphics[width=1.0\textwidth]{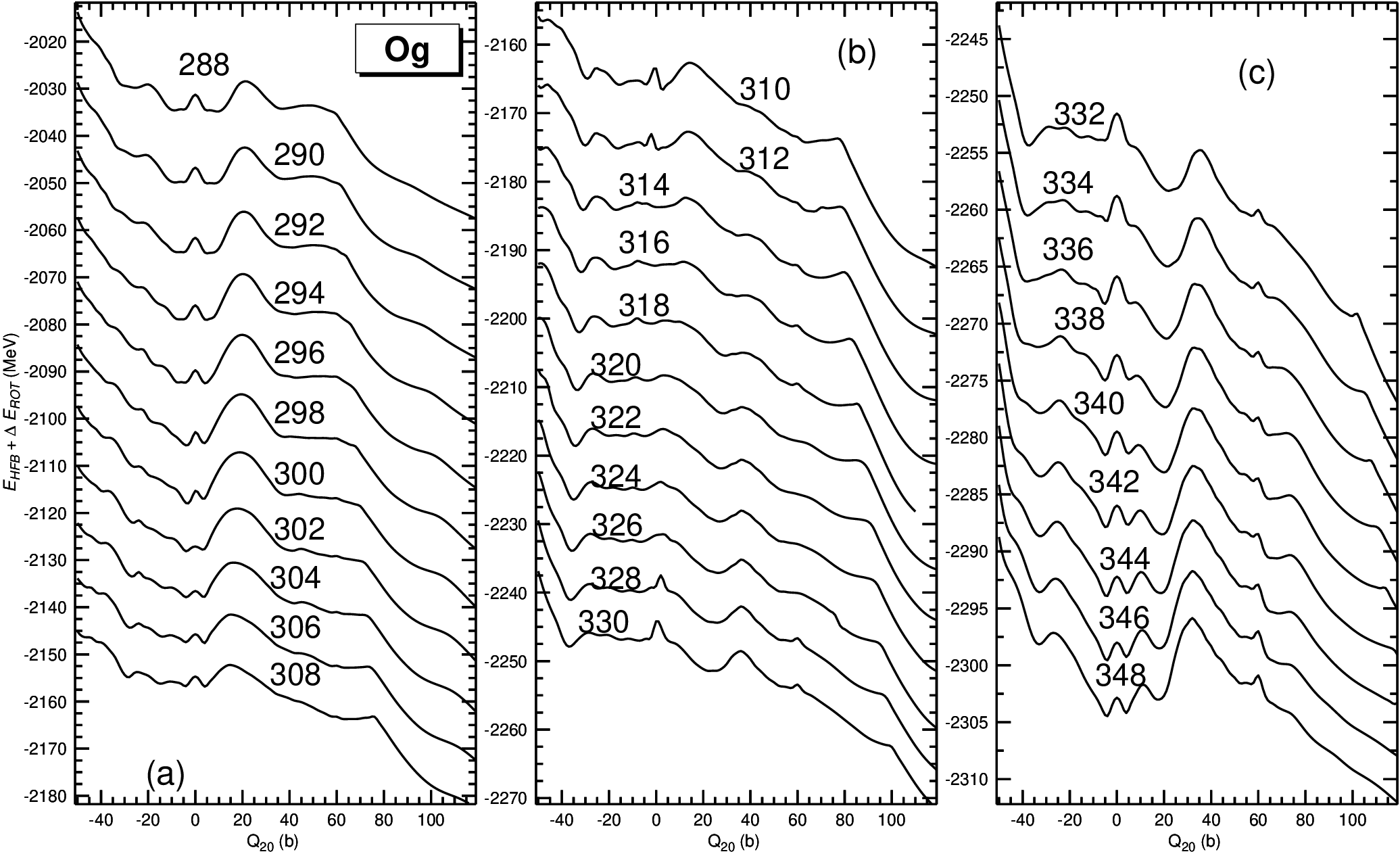} \
\caption{ 
The HFB plus the zero point rotational energies, obtained with the  
Gogny-D1M$^{*}$ EDF, are plotted  as functions of the quadrupole moment 
Q$_{20}$ for the nuclei $^{288-348}$Og. Starting from $^{290}$Og [panel 
(a)], $^{312}$Og [panel (b)] and $^{334}$Og [panel (c)], the curves 
have been successively  shifted by 2.5 MeV in order to accommodate them 
in a single plot. Note that the energy scales span different ranges in 
each panel. For more details, see the main text. 
}
\label{FissionBarriersOg} 
\end{figure*}

The two-proton separation energies S$_{2P}$ are depicted in panel (b) 
of Fig.~\ref{two-nucleon-sep-ener-FIG} as functions of the proton 
number Z. Starting with the N = 152 isotones, the curves have been 
shifted by  4 MeV. As can be seen, with few exceptions, the S$_{2P}$ 
values display a structureless behavior. Only for isotones with 186 
$\le$ N $\le$ 204, we find some irregularities  associated with sudden 
shape transitions. As with the two-neutron separation energies, a more 
realistic treatment of those irregularities requires  the inclusion of 
dynamical correlations \cite{Rayner-Sharma,global-q2-corre}. Along 
isotonic chains, the S$_{2P}$ values decrease with increasing proton 
number as one moves towards the two-proton driplines. In our 
calculations, the nuclei $^{274}$Fl, $^{280}$Lv, $^{298}$122, 
$^{304}$124, $^{306}$124 and $^{312}$126 are slightly unbound with 
S$_{2P}$ = -0.28, -0.35, -0.28, -0.83, -0.10 and -0.77 MeV, 
respectively. Our calculations agree well with the available 
experimental  data \cite{S2N-S2P-EXP} as well as with results obtained 
with the D1M parametrization. For example, for  $^{250-256}$Fm, we have 
obtained S$_{2P}$ = 8.20, 9.21, 10.11, and 10.95 MeV with D1M$^{*}$ and  
S$_{2P}$ = 7.62, 8.59, 9.47, and 10.31 MeV with D1M. The previous 
values should be compared with the experimental ones \cite{S2N-S2P-EXP} 
i.e., S$_{2P}$ = 7.74, 8.93, 9.71 and 10.43 MeV.

The comparison of the S$_{2N}$ and S$_{2P}$ values, obtained with
the Gogny-D1M$^{*}$ EDF, with the available experimental
data provides the  rms values $\sigma(\mathrm{S}_{2N})=0.36 \mathrm{MeV}$ and
$\sigma(\mathrm{S}_{2P})=0.52 \mathrm{MeV}$.

Let us perform a more detailed  description of the  fission paths of 
the studied nuclei. We stress, that reflection symmetry has been 
allowed to break at any stage of our calculations. As a consequence, 
octupole deformed shapes have been taken into account, whenever they are 
favored energetically, along the minimal energy (one-dimensional 
projected) fission paths discussed in what follows as functions of the 
driving quadrupole moment Q$_{20}$.

The  HFB plus the zero point rotational energies, are plotted in 
Fig.~\ref{FissionBarriersFm} for  $^{250-304}$Fm. The main features of 
the fission paths for 150 $\le$ N $\le$ 170 are a prolate ground state 
and a wide fission barrier. Shallow fission isomers and outer barriers 
have been obtained for the lighter isotopes but, they disappear with 
increasing N. A shoulder develops between the ground state and the  top 
of the fission barrier becoming more apparent at N =170. The isotopes 
with 172 $\le$ N $\le$ 178 are still prolate though with smaller 
deformations. Here, the shoulder becomes a deeper prolate local minimum 
and the fission barriers are two-humped.
 
For 180 $\le$ N $\le$ 192, the ground states are weakly deformed or 
spherical. In particular, $^{280-284}$Fm are predicted to be weakly 
oblate. All those nuclei, become spherical at the (intrinsic) HFB level 
as their neutron numbers are close to N = 184, which is predicted to be 
a magic neutron number in our Gogny-D1M$^{*}$ calculations as well as
with the D1M parametrization. 
The nonzero ground state Q$_{20}$ value is a direct consequence of the 
rotational energy correction that shifts spherical HFB minima to  
nonzero quadrupole moments 
\cite{RRG23S,ER-Lectures,Rayner-PRC2004,NPA-2002,Rayner-fission-1}. The 
prolate local minimum decreases its excitation energy with increasing 
N. Starting at N=186  the top of the second 
barrier develops an additional minimum and the fission paths for N$>$186 become three-humped. The prolate local minimum becomes the ground 
state for 194 $\le$ N $\le$ 204. The associated prolate wells increase 
their depth and  two-humped barriers determine the stability of these 
isotopes against fission. Similar results have been obtained for the No 
and Rf chains.

The  HFB plus the zero point rotational energies, are shown in 
Fig.~\ref{FissionBarriersSg} for  $^{256-314}$Sg. Many of the features 
obtained for the Fm, No and Rf nuclei are still present for the Sg 
isotopes. For example,  the nuclei with  150 $\le$ N $\le$ 170 are 
prolate with wide  barriers. The shallow fission isomers drive the 
existence of the outer barriers and disappear with increasing N. A 
shoulder emerges between the ground state and the top of the fission  
barrier for $^{268}$Sg and becomes more apparent for $^{276}$Sg. The 
prolate deformations are further  reduced for  172 $\le$ N $\le$ 178, 
exception made of the oblate ground state obtained for $^{282}$Sg. With 
increasing N the  shoulder becomes a prolate local minimum and the 
fission barriers are two-humped.

For 180 $\le$ N $\le$ 188, the ground states are weakly deformed or 
spherical. The prolate local minimum becomes deeper with increasing N 
and for the nuclei $^{292,294}$Sg an additional minimum is observed at the 
top of the second barrier. For 190 $\le$ N $\le$ 208 the prolate local minimum 
further decreases its energy and becomes the ground state. The outer 
sectors of the fission paths exhibit a faster decline than the ones  
for Fm, No and Rf isotones. A fission barrier with a single hump 
determines the stability of the heaviest isotopes. Similar features 
have been found for Hs and Ds nuclei.

The  HFB plus the zero point rotational energies, are depicted in 
Fig.~\ref{FissionBarriersCn} for  $^{270-332}$Cn. The main features of 
the fission paths obtained for $^{270-290}$Cn (158 $\le$ N $\le$ 178) 
are their prolate ground states and their wide two-humped  barriers. For 180 
$\le$ N $\le$ 192, the ground states become weakly deformed or 
spherical. With increasing neutron number, the tendency found in 
lighter isotopes (i.e., the decrease in the height of the second hump 
of the original  barrier) is reinforced due to the already very 
pronounced decline in the outer sectors of the fission paths. The inner barrier 
(i.e., the first hump of the original barrier) still plays a role 
though its height is  severely reduced. A prolate well starts to 
develop after the inner barrier in the N=190 nucleus $^{302}$Cn.  With 
increasing neutron number, this prolate well becomes deeper and the 
associated minimum turns out to be the ground state for 194 $\le$ N 
$\le$ 220. Moreover, the outer sectors of the fission paths display a 
pronounced decline and a barrier with a single hump is obtained for the 
heaviest isotopes. A similar structural evolution has been obtained for 
the fission paths of the nuclei $^{274-338}$Fl and $^{280-344}$Lv. 

%
\begin{figure*}
\includegraphics[width=1.0\textwidth]{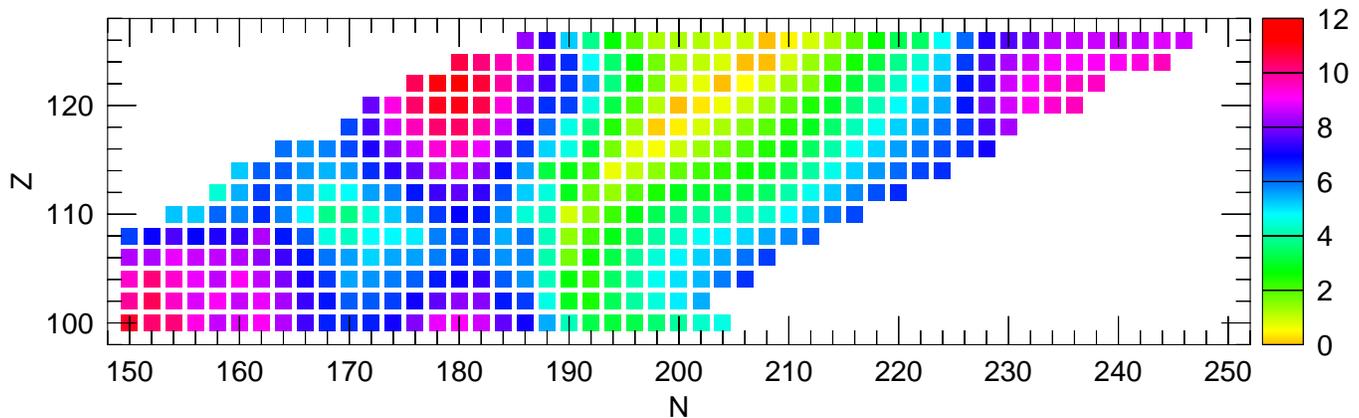} 
\caption{(Color online) The height of the largest fission barrier for 
each nucleus is shown with a color code (on the right hand side, in 
MeV). Results have been obtained with the Gogny-D1M$^{*}$ EDF.
}
\label{largest-B} 
\end{figure*}

The  HFB plus the zero point rotational energies are plotted in 
Fig.~\ref{FissionBarriersOg} for  $^{288-348}$Og. For $^{288-296}$Og 
(170 $\le$ N $\le$ 178) the ground states are prolate. Their fission 
barriers are wide with a prominent first hump followed, by a rather 
flat region. This flat region, further decreases its energy leading to 
a pronounced decline in the outer sectors of the fission paths obtained 
for $^{298-316}$Og (180 $\le$ N $\le$ 198). The original hump reduces 
its height substantially and disappears with increasing N. For these 
nuclei, we also observe a pronounced reduction in the energy of the 
oblate configurations with increasing N. As a result, some of the 
fission paths exhibit a very complex topology characterized by a strong 
competition between oblate, spherical and prolate  minima that makes 
the identification (oblate, spherical or prolate) of the actual ground 
state deformation in this kind of  neutron-rich nuclei rather involved 
\cite{Giuliani-Pinedo-SHE-rp,Warda-Egido-2012,cea_compilation}. 
Obviously, what is needed, in the presence of such a complex topology 
arising  from multiple shape coexistent minima, is to account for
quadrupole configuration mixing effects in the spirit of the GCM approach 
\cite{rs,Rayner-Sharma,global-q2-corre}. However, such a gigantic task 
is out of the scope of this study for several technical reasons (for 
example, the large number of HO shells used and the number of degrees 
of freedom required in the GCM ansatz).

For larger neutron numbers (200 $\le$ N $\le$ 230), the ground states 
are  prolate in the case of $^{318-342}$Og (200 $\le$ N $\le$ 224). The 
prolate wells (Q$_{20}$ $\approx$ 20b) become deeper   while the 
fission barriers (Q$_{20}$ $\approx$ 36b) increase their heights. With 
increasing neutron number, however, we also observe the development of  
weakly deformed minima, around Q$_{20}$ = 0. Those additional minima 
decrease their energies and already for $^{342}$Og, we observe a very 
strong competition between them and the prolate ground state. This 
multipole shape coexistence also extends to the heaviest isotopes 
$^{344-348}$Og whose ground states are predicted to be weakly oblate. 
For all these isotopes the energies of the fission paths decrease very 
quickly, as functions of the quadrupole moment, beyond the fission 
barriers. Similar comments do also apply for isotopic chains with 
larger Z values, i.e., Z = 120, 122, 124 and 126.

%

\begin{figure*}
\includegraphics[width=1.0\textwidth]{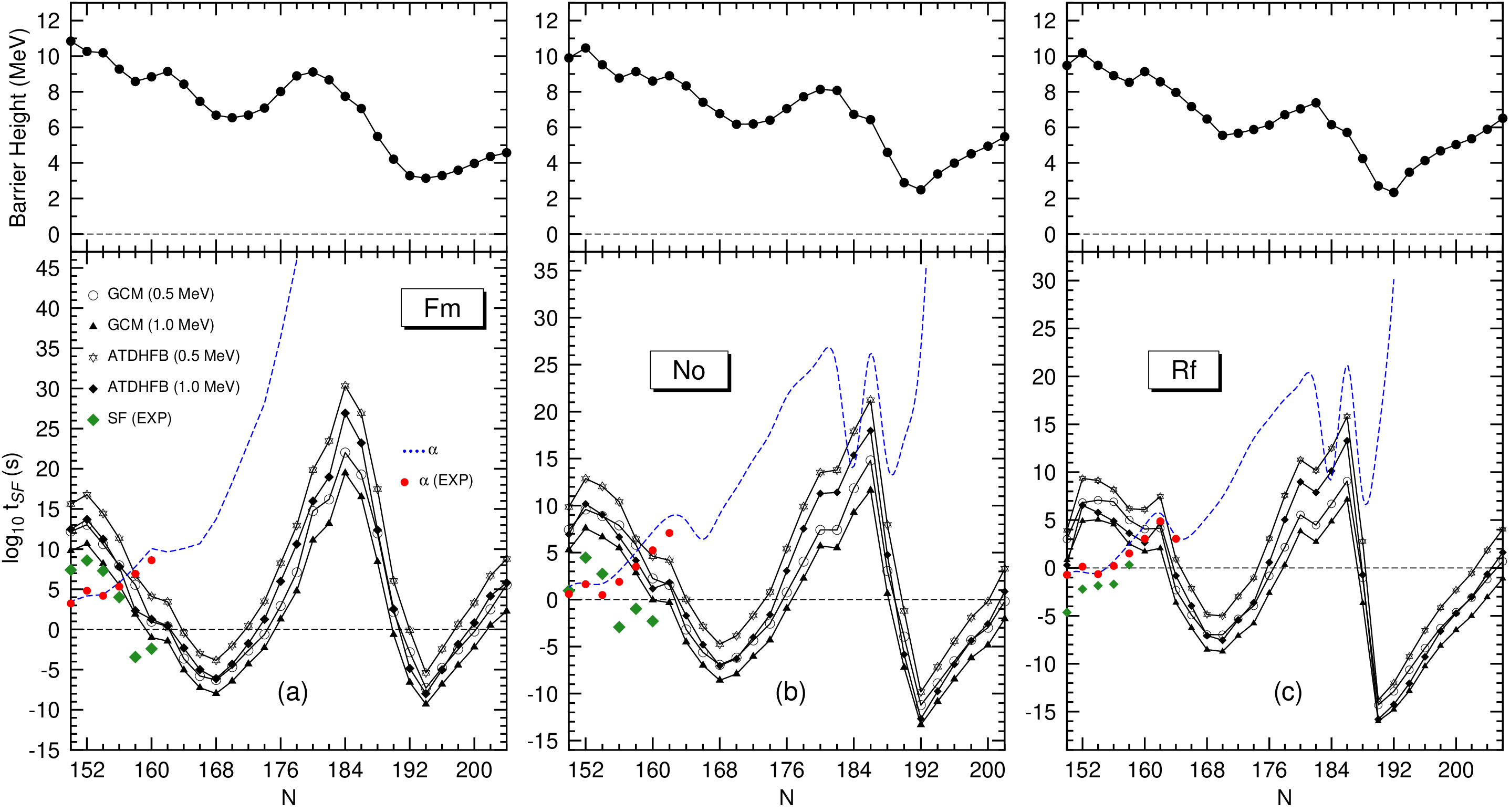} \
\includegraphics[width=1.0\textwidth]{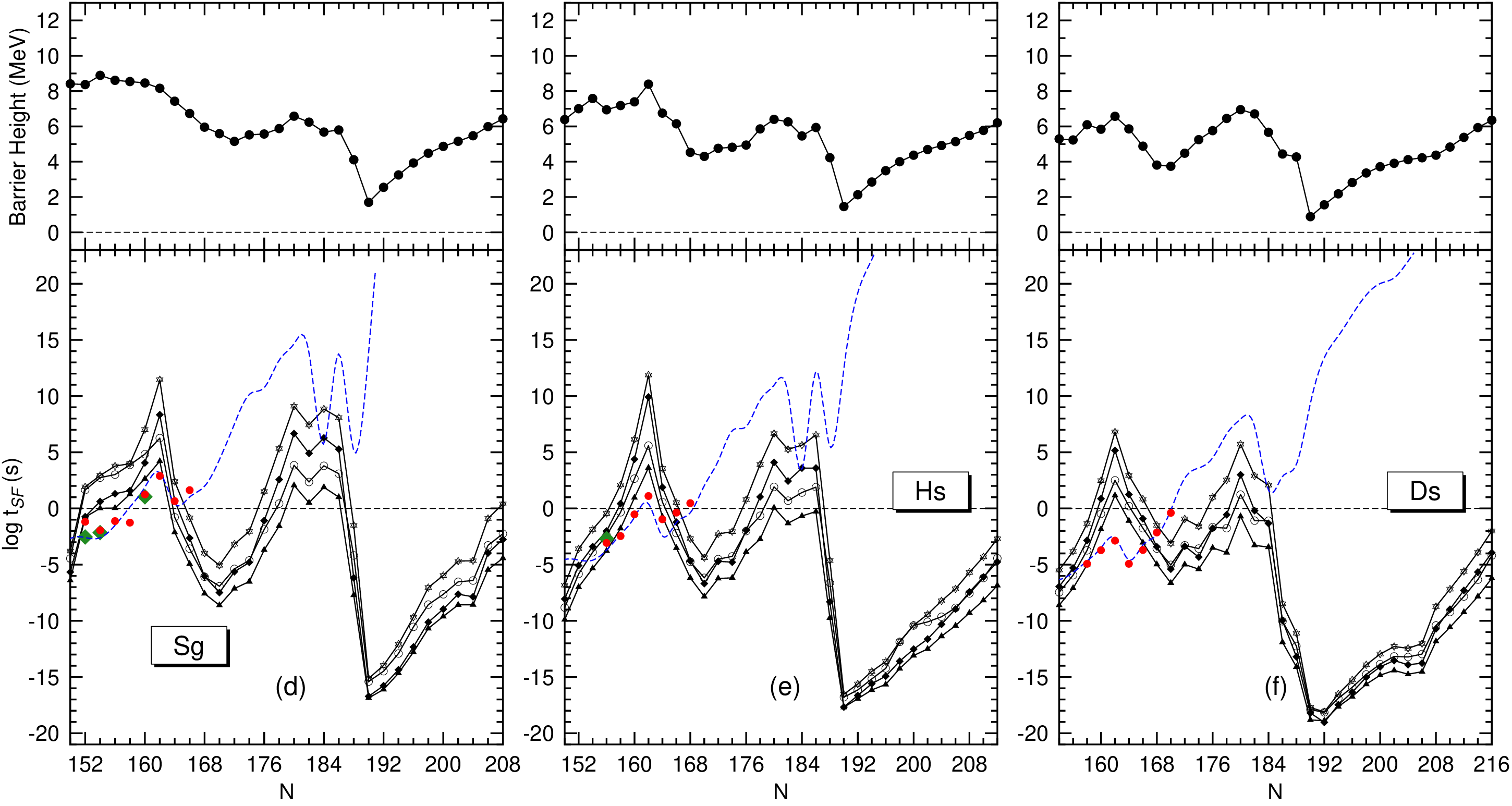} 
\caption{(Color online) The spontaneous fission half-lives 
t$_\mathrm{SF}$, predicted within the GCM and ATDHFB schemes for the 
collective inertias, are shown for the Fm, No, Rf, Sg, Hs and Ds 
isotopic chains as functions of the neutron number in panels (a) to 
(f). Results have been obtained with the Gogny-D1M$^{*}$ EDF. 
Calculations have been carried out with E$_{0}$ = 0.5 MeV and E$_{0}$ 
=1.0 MeV. The experimental t$_\mathrm{SF}$ values 
\cite{Refs-barriers-other-nuclei-3-tsf,Bertsch15} are included in the plots as 
green diamonds. The $\alpha$-decay half-lives are plotted with short 
dashed lines in each panel. Experimental values (red dots) have been 
taken from Ref.~\cite{S2N-S2P-EXP}. In addition, the largest fission 
barrier heights are plotted at the top of each panel. For more details, 
see the main text.
}
\label{tsf-part-1} 
\end{figure*}

As some of the fission barriers are rather low, we decided to check in those
cases the effect of triaxial shapes in their heights. For this purpose
we have carried out calculations in four representative nuclei, $^{300}$Hs, 
$^{306}$Cn, $^{314}$Cn and $^{310}$Fl. For those nuclei the axial barrier
heights are 2.13, 1.42, 3.06 and 1.05 MeV, respectively. The inclusion of
triaxial shapes lowers the barriers by 0.08, 0.06, 0 and 0.06 MeV, respectively.
The maximum value of the triaxial deformation parameter $\gamma$ never 
exceeds 5 degrees. Combining these results with those of Sec \ref{validate}
we observe a correlation between the height of the axial barrier and the
amount of energy gained by allowing triaxial shapes: the largest the axial
barrier height, the largest is the energy gain due to triaxial shapes. At
least in the examples studied the triaxial degree of freedom seems to play
a minor role when the axial barrier heights are less than 3 or 4 MeV.

For each nucleus, the largest barrier height
is plotted in Fig.~\ref{largest-B}
with a color code. The barriers are obtained without taking into account triaxial
shapes and therefore the values given have to be considered as upper limits. Their evolution, as functions 
of the neutron number, provides a rough estimation of the stability of 
the considered nuclei against fission. As can be seen from the figure, 
our Gogny-D1M$^{*}$ calculations predict two regions of local maxima 
around $^{250}$Fm and $^{300}$120. Those regions are usually referred 
as the "peninsula of known nuclei" and the "island of stability", 
respectively. In our calculations, we have obtained barrier heights of 
up to 10.85 MeV for the former while for the latter the predicted 
heights reach up to 11.19 MeV. In between these two regions, the 
heights exhibit minima around  N = 168-170 and decrease up to around 4 
MeV. On the other hand, the lowest values of the barrier heights 
correspond to nuclei  with neutron numbers 188 $\le$ N $\le$ 208.  In 
particular, vanishing barriers have been obtained for nuclei with the 
proton to neutron ratios Z/N = 118/198, 118/200, 120/200, 122/202, 
124/206, 124/208 and 126/208. Regions of vanishing barriers have 
already been predicted within the Mac-Mic and mean-field approaches 
\cite{Sierk-PRC2015,Giuliani-Pinedo-SHE-rp,Capote-Gor,vanishing-Agbe-2017-RMF}. They play an 
important role to determine the Z/N ratios around which the r-process 
nucleosynthesis of superheavy elements is terminated by neutron induced 
fission \cite{terminated-1,terminated-2}. Beyond those regions of 
vanishing barriers, the heights increase as functions of N for all the 
considered isotopic chains. Qualitatively, our results compare well 
with the ones obtained in previous studies 
\cite{Sierk-PRC2015,Giuliani-Pinedo-SHE-rp,Capote-Gor,vanishing-Agbe-2017-RMF} though  
quantitative differences are observed, especially for neutron-rich 
nuclei. However, calculations of fission reaction rates are required to 
better understand the sensitivity of the r-process abundances with 
respect to the fission properties predicted within different 
theoretical models.

\begin{figure*}
\includegraphics[width=1.0\textwidth]{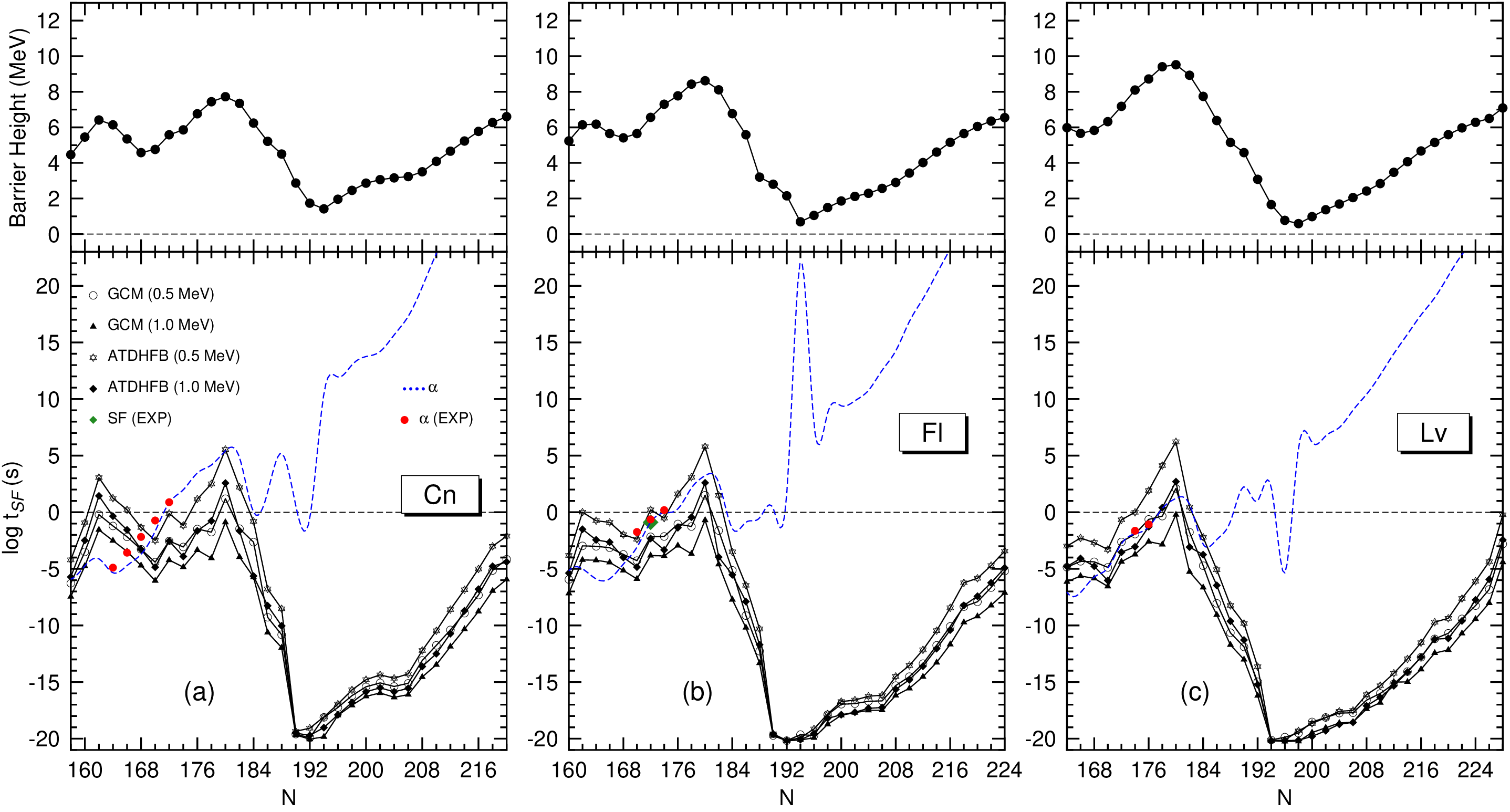} \
\includegraphics[width=1.0\textwidth]{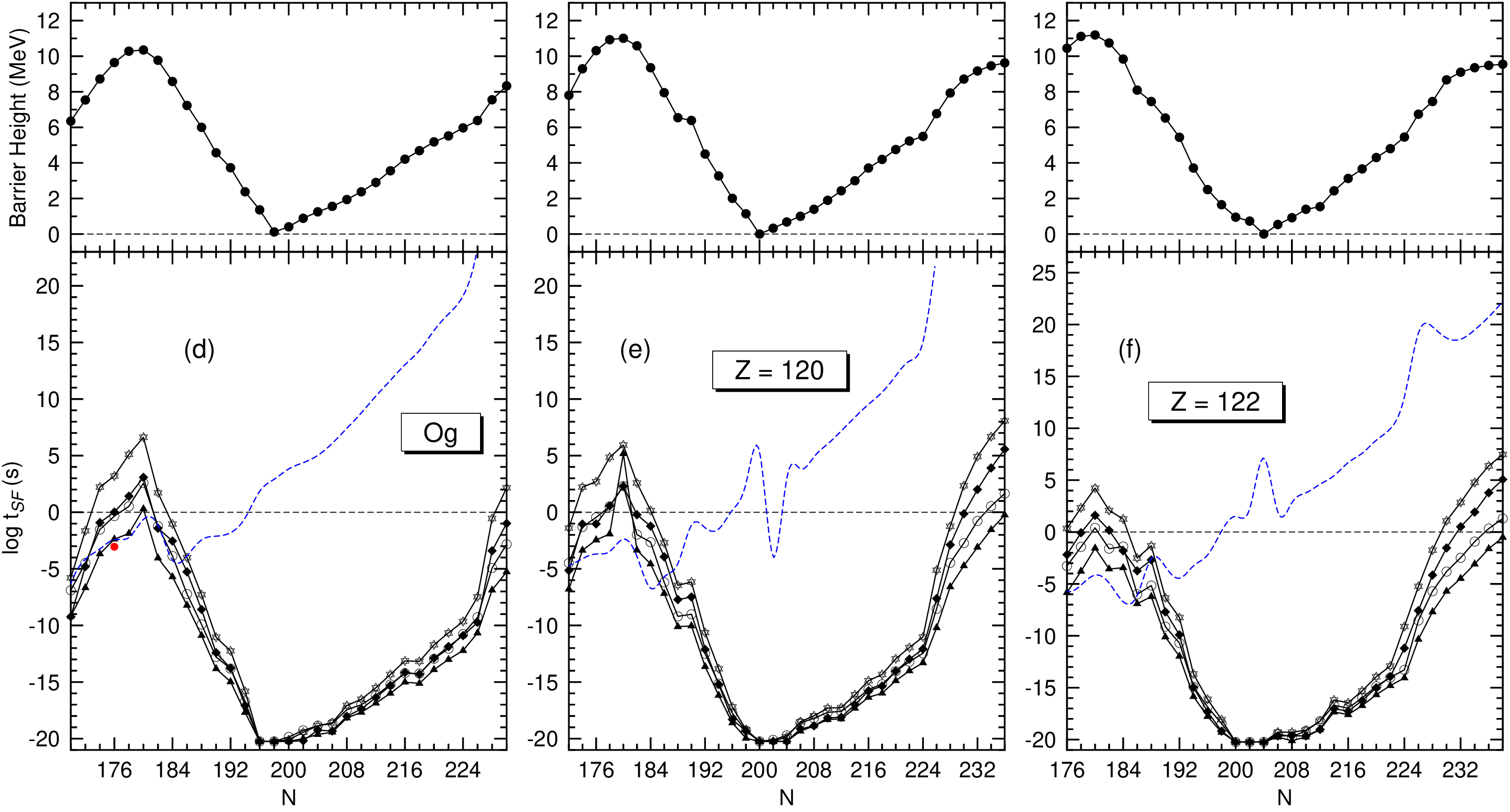} 
\caption{(Color online) The same as Fig.~\ref{tsf-part-1} 
but for the Cn, Fl, Lv, Og, Z =120 and  Z = 122  isotopic chains.
}
\label{tsf-part-2} 
\end{figure*}

\begin{figure*}
\includegraphics[width=1.0\textwidth]{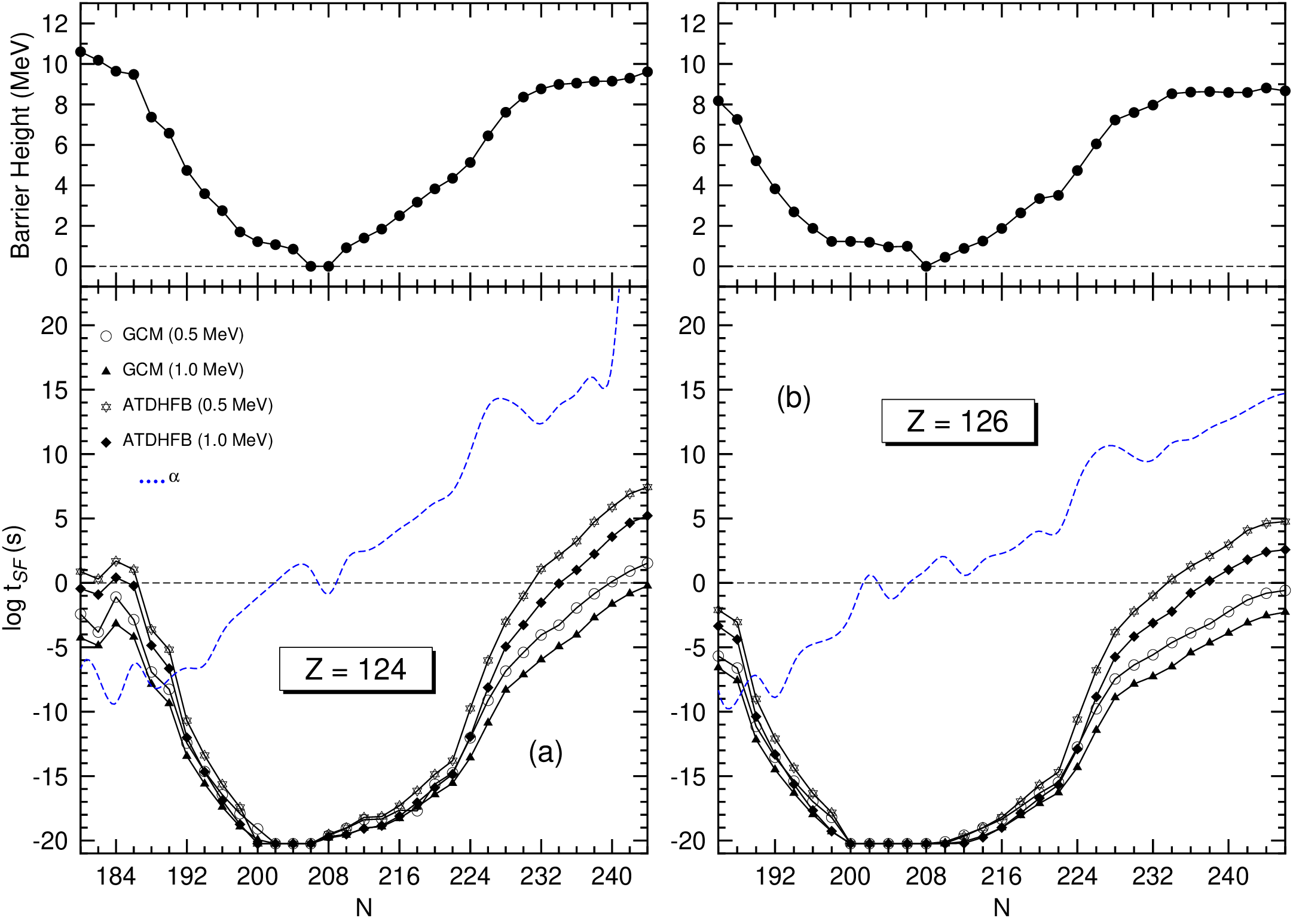} 
\caption{(Color online) The same as Fig.~\ref{tsf-part-1} 
but for Z =124 and  Z = 126  isotopic chains.
}
\label{tsf-part-3} 
\end{figure*}

The spontaneous fission half-lives t$_\mathrm{SF}$, predicted within 
the GCM and ATDHFB schemes for the collective inertias, are shown in 
panels (a) to (f) of Fig.~\ref{tsf-part-1} for the Fm, No, Rf, Sg, Hs 
and Ds isotopic chains as functions of the neutron number. The 
corresponding results  for the Cn, Fl, Lv, Og, Z =120 and  Z = 122 
chains are shown in Fig.~\ref{tsf-part-2} and the ones for nuclei with 
Z = 124 and Z =126 in Fig.~\ref{tsf-part-3}. The largest barrier height
(see, Fig.~\ref{largest-B}) is plotted at the top of 
each panel for each isotope. 

Before discussing the t$_{\mathrm{SF}}$ values let us mention an 
assumption made to deal with those nuclei with an oblate ground state. 
In this case, the nucleus can proceed to fission by climbing the 
spherical maximum or it can take a detour through the $\gamma$ plane 
(see \cite{Ryssens.19} for a recent example).  
As the energy of the oblate minimum is, in most of the cases, rather 
close to the one of the prolate configuration we are in the typical 
case of shape coexistence. In those cases, the oblate and prolate 
minima have roughly the same absolute value of $\beta_{2}$ and are 
connected through an arc in the $\gamma$ plane. As a systematic 
calculation of triaxial configurations is prohibitive for the large 
configuration spaces used, we have assumed that the contribution to the 
action (entering the WKB formula) from the triaxial path connecting the 
oblate and prolate minimum or from the the axially symmetric path going  
through the spherical maximum can be neglected. This assumption is 
justified by the expected very small energy barrier between the oblate 
and prolate minima that lead to a negligible contribution to the 
action, as compared to the one corresponding to the path going from the 
prolate minimum to scission through the first fission barrier, as this 
barrier is usually far much larger \cite{triaxial-cont}  (see, for 
instance, the case of $^{300}$Og and neighboring nuclei, in Fig. 
\ref{FissionBarriersOg}).

Regardless of the E$_{0}$ value and/or the  scheme employed, the trend 
in the predicted spontaneous fission half-lives resembles, as expected, 
the one obtained for the barrier heights. Nevertheless, for a given 
nucleus, the probability to penetrate the fission barrier depends on 
several other ingredients and cannot be solely determined by the 
barrier height. In particular, one should keep in mind that the 
predicted t$_\mathrm{SF}$ values are also affected, for example, by the 
barrier shapes (mainly their width) and the behavior and size of the 
collective inertia. Therefore, calculations involving the latter should 
be carried out to make quantitative predictions. 

%

For all the studied isotopic chains, the spontaneous fission half-lives 
display maxima around N = 180-186, that reflect the effect of the N = 
184 neutron shell closure. Immediately after those maxima there is a 
sudden dip around N = 188-194 for 100 $\le$ Z $\le$ 118 and around N = 
200 for 120 $\le$ Z $\le$ 126 with very small values of 
t$_\mathrm{SF}$. Those very small values (up to t$_\mathrm{SF}$ 
$\approx$ 10$^{-21}$ s) correspond to nuclei with very small and/or 
vanishing barrier heights. The subsequent increase observed in the 
t$_\mathrm{SF}$ values for larger neutron numbers correlates well with 
the increase observed in the heights of the (prolate) barriers that 
emerge in the fission paths of the heaviest nuclei in each isotopic 
chain (see, Figs.~\ref{FissionBarriersFm}, \ref{FissionBarriersSg}, 
\ref{FissionBarriersCn}, \ref{FissionBarriersOg}). For  Fm, No and Rf 
nuclei, our calculations predict local maxima at N = 152  that reflect 
shell effects arising from the corresponding gap in the neutron 
single-particle spectrum [see, panel (a) of  Fig.~\ref{spe_266Rf}]. In 
the case of Fm and No isotopes, a small kink is observed around N = 162 
whereas for Rf nuclei the effect of this deformed shell closure on the 
lifetimes is more pronounced. On the other hand, for 106 $\le$ Z $\le$ 
114 the first  local maximum is located at  N = 162. Note that, with 
increasing Z values, we observe a shape change in the fission barriers 
around N = 162. First, for low Z values a single barrier is obtained 
around N = 162. Then, the barrier becomes wider and two-humped. 
However, further increasing Z we observe a severe reduction in the 
height of the second hump together with a pronounced decline in the 
outer sections of the fission paths. This leads to  much smaller 
t$_\mathrm{SF}$ values for Fl isotopes around N = 162 as compared with, 
for example, the ones obtained for Cn isotopes around the same neutron 
number. In our Gogny-D1M$^{*}$ calculations, both the barrier heights 
and the spontaneous fission half-lives exhibit a local minimum around N 
= 168-170 which also correlates well with the structural evolution 
observed in the fission paths around these neutron numbers.

Although the previous discussion shows that the trend with neutron 
number of t$_\mathrm{SF}$ is rather insensitive to the details, like 
the E$_{0}$ value used or the scheme adopted for the calculation of the 
collective inertias, the specific values of t$_\mathrm{SF}$ really 
depend on them. For instance, the lifetimes obtained with the ATDHFB 
inertias tend to be larger than the ones obtained in the GCM framework 
for a given E$_{0}$. The difference becomes more pronounced the higher 
and wider the fission barriers are as can be observed in previous 
calculations for neutron-rich Ra, U and Pu nuclei 
\cite{Rayner-fission-1,Rayner-fission-2,Rayner-fission-3}. For example, 
we have obtained for the lighter Fm isotopes $^{250-254}$Fm, with  
E$_{0}$ = 0.5 MeV, the values $\log_{10}$ t$_\mathrm{SF}^{GCM}$ = 
12.19, 13.03, 10.62 and $\log_{10}$ t$_\mathrm{SF}^{ATDHFB}$ = 15.60, 
16.77, 14.44. The comparison with the available experimental 
t$_\mathrm{SF}$ values 
\cite{Refs-barriers-other-nuclei-3-tsf,Bertsch15} reveals that  our 
results overestimate them though the trend with neutron number is 
reproduced reasonably well in most of the cases. Larger E$_{0}$ values 
are required to improve the agreement with the experiment. For example, 
for $^{250-254}$Fm [see, panel (a) of Fig.~\ref{tsf-part-1}], with  
E$_{0}$ = 1.0 MeV, we have obtained $\log_{10}$ t$_\mathrm{SF}^{GCM}$ = 
9.81, 10.65, 8.18 and $\log_{10}$ t$_\mathrm{SF}^{ATDHFB}$ = 12.47, 
13.68, 11.22. 

%

\begin{figure*}
\includegraphics[width=1.0\textwidth]{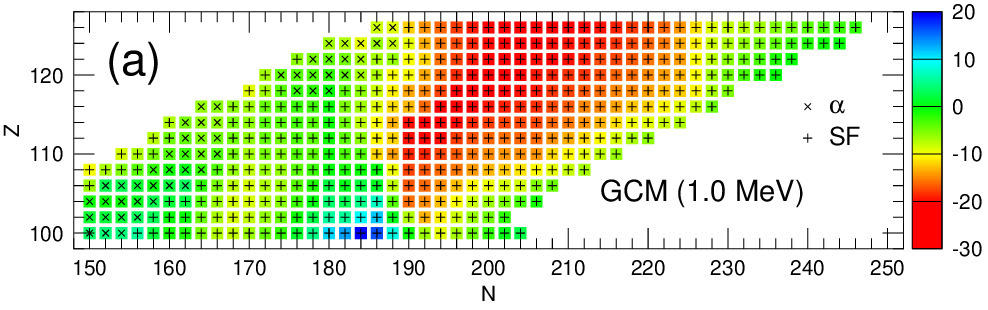} \
\includegraphics[width=1.0\textwidth]{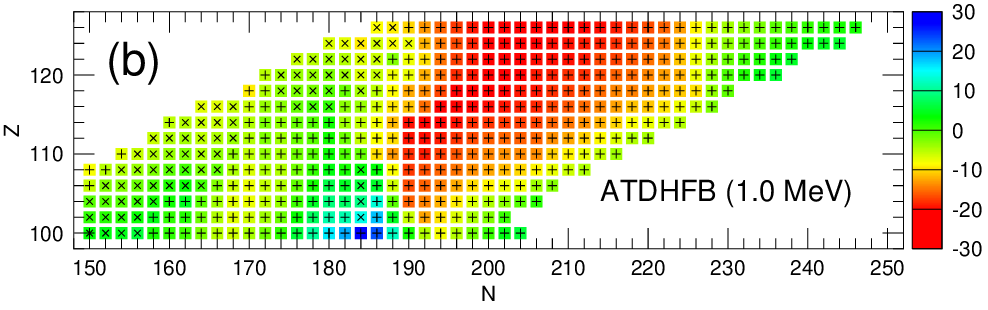} \
\caption{(Color online) Logarithm  of the shortest (spontaneous fission 
or $\alpha$-decay) half-life (is seconds) obtained for the nuclei 
studied in this work within the GCM [panel (a)] and ATDHFB [panels (b)] 
collective inertia schemes. Results correspond to the Gogny-D1M$^{*}$ 
EDF and E$_{0}$ = 1.0 MeV. For more details, see the main text.
}
\label{domin-decay} 
\end{figure*}

In Figs.~\ref{tsf-part-1}, \ref{tsf-part-2} and \ref{tsf-part-3}, we 
have also plotted the $\alpha$-decay half-lives computed with  a 
parametrization \cite{TDong2005} of the  Viola-Seaborg formula 
\cite{Viola-Seaborg} Eq. (\ref{VSeaborg-new}). The predicted 
$\alpha$-decay lifetimes agree well with the available experimental 
data \cite{S2N-S2P-EXP}. A comparison of the $Q_{\alpha}$ values 
obtained in our Gogny-D1M$^{*}$ calculations with the available 
experimental data yields the rms value  $\sigma (Q_{\alpha})=0.27$ MeV. 
The same comparison for $\log_{10}t_{\alpha}$ yields $\sigma ( 
\log_{10}t_{\alpha}) =0.84$. On the one hand, $\alpha$-decay mainly 
dominates in the proton-rich regions. On the other hand, fission turns 
out to be faster than $\alpha$-decay with increasing neutron number. In 
Fig.~\ref{domin-decay}, we have plotted the logarithm of the shortest 
half-life (in seconds) for the studied nuclei. In spite of the 
quantitative differences, the patterns emerging within the GCM [panel 
(a)] and ATDHFB [panel (b)] schemes for the collective inertias, with 
E$_{0}$ = 1.0 MeV, are very similar. The same is also true for E$_{0}$ 
= 0.5 MeV and, therefore, they are not shown in the figure. Typically, 
experimental techniques allow the full characterization of nuclei with 
half-lives larger than 10 $\mu$s. In our calculations, long living 
nuclei are predicted around Z = 106 and N = 160. For example, for 
$^{266}$Sg we have obtained half-lives around 10$^{2}$ s. A second 
region of long living nuclei is predicted around N = 180. For example, 
for the N = 180 isotones  $^{288}$Hs, $^{290}$Ds, $^{292}$Cn and 
$^{294}$Fl, for which fission is the dominant decay mode, we have 
obtained t$_\mathrm{SF}^{GCM}$ = 1.16, 0.19, 0.13 and  0.20 s  while 
t$_\mathrm{SF}^{ATDHFB}$ $\approx$ 10$^{4}$ s for $^{288}$Hs and 
t$_\mathrm{SF}^{ATDHFB}$ $\approx$ 10$^{3}$ s for $^{290}$Ds, 
$^{292}$Cn and $^{294}$Fl, respectively. For larger neutron numbers, we 
have obtained a region centered around Z = 120 and N = 204 for which 
the half-lives are too short to be characterized experimentally. 
Finally, the increase observed in the fission barriers obtained for the 
most neutron-rich isotopes in each of the studied chains leads to 
larger half-lives.  We find, that the overall pattern emerging from 
Fig.~\ref{domin-decay} agrees well with the ones obtained in previous 
studies \cite{Warda-Egido-2012,Giuliani-Pinedo-SHE-rp}.

%
 
%
%
\section{Summary and conclusions}
\label{conclusions}
In this work, the constrained HFB approximation, based on the new 
parametrization D1M$^{*}$ of the Gogny-EDF, is used to study the 
properties of 435 even-even superheavy elements. Besides the usual 
constrains on both the proton $\hat{Z}$ and neutron $\hat{N}$ number 
operators, we have considered constrains on the quadrupole 
$\hat{Q}_{20}$ and octupole $\hat{Q}_{30}$  as well as on the 
$\hat{Q}_{10}$ operators. Due to the very time demanding nature of the 
calculations, axial symmetry has been kept as a selfconsistent 
symmetry. A large and optimized (deformed) axially symmetric HO basis 
has been employed. Zero-point quantum corrections have always been 
added to the HFB energies  {\it{a posteriori}}. In particular, the 
rotational energy correction has been computed in terms of the Yoccoz 
moment of inertia while the GCM and ATDHFB schemes have been used to 
compute both the collective masses and vibrational energy corrections. 
Having extracted all the required building blocks from the mean-field 
calculations, we have computed the spontaneous fission half-lives 
t$_\mathrm{SF}$ within the WKB formalism. The $\alpha$-decay half-lives 
have been obtained with a parametrization of the Viola-Seaborg formula.

The newly proposed D1M$^{*}$ parametrization is based on the previous 
D1M parametrization of the Gogny force and it imposes a more realistic 
value of the slope $L$ of the symmetry energy. As a consequence, the 
properties of very neutron-rich nuclei are expected to differ from 
those of D1M and to be more accurate in the description of nuclei with 
strong proton-neutron asymmetry. Neutron rich nuclei are specially 
relevant for the description of the nucleo-synthesis mechanism. 

The  parametrization D1M$^{*}$ has been benchmarked against available 
experimental data on inner and  second barrier heights, excitation 
energies of  fission isomers and spontaneous fission half-lives in a 
selected set of Pu, Cm, Cf, Fm, No, Rf, Sg, Hs and Fl nuclei. It is 
shown that the global trends observed in the experiment are reproduced 
reasonably well. The second barrier heights come up a bit too high (2-4 MeV),
a discrepancy that is not expected to be reduced by including triaxial shapes.
This disagreement with the experimental data remains to be analyzed in detail.

We have described the methodology employed to obtain the fission paths 
of the studied nuclei and paid special attention to proton and neutron 
pairing correlations, octupole and hexadecupole deformations as well as 
collective GCM and ATDHFB masses. Energy gaps in the proton and neutron 
single-particle spectra have been analyzed. From the comparison with 
the available experimental data and the results obtained in previous 
studies, especially in the case of the Gogny-D1M EDF, we concluded that  
D1M$^{*}$ represents a reasonable starting point to describe fission in 
heavy and superheavy nuclei. 
  
We have  performed a detailed study of the (minimal energy) fission 
paths in  superheavy nuclei. First, we have considered the systematic 
of the ground state quadrupole $\beta_{2}$ and octupole $\beta_{3}$ 
deformation parameters. Octupole deformed ground states have been 
predicted for  nuclei with neutron numbers 186 $\le$ N $\le$ 194. 
Ground state proton and neutron pairing energies as well as two-nucleon 
separation energies have been thoroughly discussed. The latter compare 
well with the available experimental data and with results obtained 
with the Gogny-D1M EDF. 

A detailed discussion of the structural changes observed in the fission 
paths of the considered nuclei has been carried out as a function of 
both Z and N. The (largest) barrier heights resulting from those 
structural changes exhibit local maxima in the "peninsula of known 
nuclei" (i.e., around Z = 100 and N = 150) and the "island of 
stability" (i.e., around Z = 120 and N = 180) whereas local minima are 
found for nuclei with N = 168-170. Moreover, small and/or vanishing 
barrier heights have been obtained for 188 $\le$ N $\le$ 208 followed 
by a region of increasing barrier heights for all the studied isotopic 
chains. Qualitatively, our results compare well with the ones obtained 
in previous studies though  quantitative differences are observed, 
especially for neutron-rich nuclei.

The GCM and/or ATDHFB spontaneous fission half-lives, computed with 
E$_{0}$ = 0.5 MeV and E$_{0}$ = 1.0 MeV, exhibit a trend with neutron 
number that resembles the one obtained for the largest of the barrier 
heights in each isotope. However, effects associated, for example, with 
the shape and width of the fission barriers, shell closures as well as 
the collective masses also play a role to determine local variations in 
the stability against spontaneous fission. The predicted  
t$_\mathrm{SF}$ values overestimate the experimental ones though the 
trend with neutron number is reproduced reasonably well. On the other 
hand, the $\alpha$-decay half-lives obtained in our calculations agree 
well with the available experimental data. The $\alpha$-decay mainly 
dominates in the proton-rich regions of the superheavy landscape while, 
with increasing neutron number, fission becomes faster. Long living 
nuclei, with half-lives greater than 10 $\mu$s, have been found around 
Z = 106 and N = 160 and around the neutron number N =180. On the other 
hand, the half-lives for nuclei around  Z = 120 and N = 204 are too 
short to be characterized experimentally. Larger half-lives, associated 
with the increase in the corresponding barrier heights, have been 
obtained  for the most neutron-rich isotopes in each of the studied 
isotopic chains. This overall pattern,  also agrees well with the ones 
obtained in previous studies.

A long list of tasks remains to be undertaken in future studies.  For 
example, consider triaxial deformation in all the stages of the
calculation to justify or discredit the general arguments used to skip its
impact on the evaluation of $t_\mathrm{SF}$. Calculations for odd-even and odd-odd superheavy nuclei 
should be carried out in order to further assess the predictive 
power of the Gogny-D1M$^{*}$ EDF. Given the technical difficulties 
associated with those calculations, the Equal Filling Approximation 
(EFA) appears as a plausible (selfconsistent) candidate for such 
studies \cite{Rayner-fission-4,Rayner-fission-5}. Moreover, it has been 
recently shown \cite{Min_Action_GognyRayner2018} that for Fm and No 
nuclei the inclusion of pairing fluctuations, associated with the 
spontaneously broken U(1)  particle number symmetry, within a least 
action approach  improves the agreement between the predicted 
t$_\mathrm{SF}$ values and the experiment. Within such an approach 
\cite{Min_Action_Gogny,Min_Action_GognyRayner2018}, a minimization of 
the action is carried out, with respect to some parameter correlated 
with the amount of pairing correlations, for each of the configurations 
along the minimal energy fission path of a given nucleus. Therefore, 
the (minimal energy) fission paths computed in this work might 
represent the basis for the corresponding large scale least action 
calculations. Work along these lines is in progress and will be 
reported elsewhere.

\begin{acknowledgments}
The work of LMR was partly supported by the Spanish
MINECO Grant No.~FPA2015-65929, No.~FIS2015-63770 and No. PGC2018-094583-B-I00.
\end{acknowledgments}

\end{document}